\documentstyle[11pt]{ioplppt}
\eqnobysec

\def\be{\begin{equation}}
\def\ee{\end{equation}}
\def\bd{\begin{displaymath}}
\def\ed{\end{displaymath}}
\newcommand{\ba}{\begin{eqnarray}}
\newcommand{\ea}{\end {eqnarray}}
\newcommand{\nn}{\nonumber}
\newcommand{\ket}[1]{|{#1}\rangle}
\newcommand{\bra}[1]{\langle {#1}|}

\def\BCT{\,\hbox{\hbox to -3pt{\vrule height 6.5pt width .2pt\hss}\rm
C}}
\def\BRT{\,\hbox{\hbox to -1pt{\vrule height 7.4pt width .2pt\hss}\rm
R}}
\def \eins{{\rm 1 \kern-2.8pt I }}

\def\bra{\langle}
\def\ket{\rangle}

\def\s{\sigma}
\def\sp{\sigma^+}
\def\sm{\sigma^-}

\def\rap{r_{\alpha}^+}
\def\ram{r_{\alpha}^-}
\def\rbp{r_{\beta}^+}
\def\rbm{r_{\beta}^-}
\def\psit{\tilde{\psi}}
\def\phit{\tilde{\phi}}

\def\half{\frac{1}{2}}

\def\g{\alpha_{+}}
\def\a{\alpha_{-}}
\def\k{\alpha_{z}}
\def\az{\alpha_{z}}
\def\b{\beta_{+}}
\def\d{\beta_{-}}
\def\rh{\beta_{z}}
\def\bz{\beta_{z}}
\def\BCT{\,\hbox{\hbox to -3pt{\vrule height 6.5pt width .2pt\hss}\rm
C}}
\def\BRT{\,\hbox{\hbox to -1pt{\vrule height 7.4pt width .2pt\hss}\rm
R}}
\def\bra{\langle}
\def\ket{\rangle}
\def\re{\mbox{Re}}

\def\vp{\varphi}
\def\vpd{\overline{\varphi}}
\def\be{\begin{equation}}
\def\ee{\end{equation}}
\def\psit{\tilde{\psi}}
\def\phit{\tilde{\phi}}

\begin{document}
\pagestyle{plain}
\thispagestyle{empty}
\vspace* {10mm}
\begin{center}
\large
{\bf 
The XX--model with boundaries\\
Part I: Diagonalization of the finite chain}
\end{center}
\begin{center}
\normalsize
       Ulrich Bilstein\footnote{bilstein@theoa1.physik.uni-bonn.de} and
       Birgit Wehefritz\footnote{birgit@theoa1.physik.uni-bonn.de}
        \\[1cm]
    {\it Universit\"{a}t Bonn,
                    Physikalisches Institut \\ Nu\ss allee 12,
                    D-53115 Bonn, Germany}\\[14mm]
\end{center}
{\bf Abstract:}
This is the first of three papers dealing with the XX finite quantum chain 
with arbitrary, not necessarily hermitian, boundary terms. This 
extends previous work where the 
periodic or diagonal boundary terms were considered. In order to find the 
spectrum and wave-functions an auxiliary quantum chain is examined which 
is quadratic in fermionic creation and annihilation operators and hence 
diagonalizable. The secular equation is in general complicated but 
several cases were found when it can be solved analytically. For these 
cases the ground-state energies are given. The appearance of boundary 
states is also discussed and in view to the applications considered in 
the next papers, the one and two-point functions are expressed in terms 
of Pfaffians.
\small
\noindent

\rule{5cm}{0.2mm}

\vspace{1cm}
\begin{flushleft}
\parbox[t]{3.5cm}{\bf PACS numbers:}
\parbox[t]{12.5cm}{75.10.Jm, 05.50.+q}
\end{flushleft}
\normalsize

\newpage
\pagestyle{plain}
\setcounter{page}{1}

\section{Introduction}

In this paper, we consider the 
XX--chain with diagonal and non--diagonal boundary terms:
\be
\label{Ham}
\fl
H = \frac{1}{2} \;\sum_{j=1}^{L-1} \Bigl[ \sp_j \sm_{j+1} +
\sm_j \sp_{j+1} \Bigr]+
\frac{1}{\sqrt{8}} \Bigl[ \a \sm_1 + \g \sp_1 +\k \s^{z}_1
+ \b \sp_L + \d \sm_L +  \rh \s^{z}_L \Bigr]
\ee
Here, $\sigma^{\pm}$ are defined 
by $\sigma^{\pm}=\frac{1}{2}(\sigma^x \pm \i\sigma^y)$, 
where $\sigma^x, \sigma^y$ and $\sigma^z$ are the Pauli matrices.
The factor $\frac{1}{\sqrt{8}}$ has been introduced for later convenience.
Since the parameters $\alpha_{\pm},\beta_{\pm},\az$ and $\bz$ are
arbitrary complex numbers,
the Hamiltonian defined by eq.\ (\ref{Ham}) is non--hermitian in the 
general case.

Let us now give a brief overview of the literature
before turning to the concrete results we obtained by studying
diagonal and non--diagonal boundary conditions.
All the articles mentioned in this overview
are based on the free fermion approach to the XX--model. 

The XX--model often appears as a special case
of the XY--model. The XY--model was introduced 1961 by 
Lieb, Schultz and Mattis \cite{LSM} who computed  its ground
state energy, the elementary excitations and also
presented a method to calculate  time--independent
correlation functions. In this way,
they treated periodic boundary conditions as well as free ends.

During the last thirty years, the correlation functions of 
the XY--model and therewith the XX--model have been
the subject of various investigations.
McCoy \cite{McCoy} studied the correlation functions of the XY--model
with periodic boundary conditions. More precisely, he computed
the asymptotic behaviour of each of the
three time--independent correlation functions 
$\langle \sigma_0^i\sigma_R^i\rangle$ with $i=x,y,z$ in the 
limit $R \rightarrow \infty$. Barouch and McCoy \cite{Barouch_McCoy}
determined the asymptotic behaviour of the same correlation functions
for the XY--model with an
external time--independent magnetic field in the $z$--direction. 
In another article, time-dependent spin--spin correlation functions of the form 
$\langle \sigma_0^x(t)\sigma_R^x(0)\rangle$ and
$\langle \sigma_0^y(t)\sigma_R^y(0)\rangle$ for the 
XY--model in an external magnetic field again in the $z$--direction 
were calculated
in the limit of large $R$
by McCoy, Barouch and Abraham \cite{MBA}. Exact expressions for these
correlation functions for all values of $R$
and $t$ were then computed by Vaida and Tracy \cite{Vaida_Tracy}.
Furthermore, time--dependent 
many--spin correlation functions for the XY--model in an external
constant magnetic field in the $z$--direction were treated 
by Bariev \cite{Bariev}.

Recently, the XY--model with boundary terms has been a subject 
of rising interest. Hinrichsen and Rittenberg \cite{Vladimir_Haye} 
showed that the anisotropic XY--model
in an external magnetic field with $\sigma^z$--boundary terms 
is invariant under certain quantum group transformations. Furthermore, they
defined and calculated the corresponding invariant correlation functions. 

The XX--model with non--diagonal boundary terms, however, has not been studied
thoroughly before. Some work in this direction has been presented 
by Guinea \cite{Guinea} who studied the semi--infinite XY--model with 
one $\sigma^x$--boundary term (i.\ e.\ $\a = \g = 1, \; \b=\d = \az = \bz =0$).
We will mention more details of that paper
when discussing the physical applications of the Hamiltonian $H$ of
eq.\ (\ref{Ham}).
Furthermore, a study of the totally asymmetric $XX$--model with  bulk terms
of the form $\sp_j \sm_{j+1}$ and  with boundary parameters
given by
$\a  \neq 0, \b \neq 0, \g = \d = \k = \rh = 0$ in the notation of
eq.\ (\ref{Ham}) can be found in our previous paper \cite{toy_paper}. 

As already mentioned in the beginning, in the general case,
the Hamiltonian given by eq.\ (\ref{Ham}) 
is non--hermitian. Interesting physical problems involving non--hermitian
Hamiltonians can be found in several articles treating non--hermitian
quantum mechanics \cite{non-hermitian}.

\vspace{1cm}

The Hamiltonian given by eq. (\ref{Ham}) can also be used in the study
of asymmetric bulk terms.
More precisely, starting from a Hamiltonian of the
form
\be
\label{Ham_nh}
\fl
\tilde{H}= \sum_{j=1}^{L-1} \Bigl[ p  \sp_j \sm_{j+1} +
 q \sm_j \sp_{j+1} \Bigr] +
\frac{1}{\sqrt{8}} \Bigl[
\a^{\prime}\,\sigma_1^- + \g^{\prime}\,
\sp_1+\k\, \s^{z}_1
+\b^{\prime}\,\sigma_L^+  + \d^{\prime}\,
\sm_L  +\rh \,\s^{z}_L \Bigr]
\ee
one can use a similarity
transformation (see for example \cite{Fabian_Valdimir})
that transforms the asymmetric bulk terms depending on the two
parameters $p$ and $q$ of the Hamiltonian given by eq.\ 
(\ref{Ham_nh})
into symmetric bulk terms. It is convenient to choose $\sqrt{pq}=\frac{1}{2}$. 
Note that the similarity transformation
changes the boundary terms.
The corresponding transformed boundary parameters $\a, \g, \b$ and 
$\d$ of the Hamiltonian given by eq.\ (\ref{Ham}) are $L$-dependent 
now and have the expressions
\be
\label{trans}
\a =  Q^{\frac{1-L}{2}} \a^{\prime}, \;\;\;\g =  Q^{\frac{L-1}{2}} \g^{\prime},
\;\;\;
\b = Q^{\frac{1-L}{2}}\b^{\prime}, \;\;\;  \d =Q^{\frac{L-1}{2}} \d^{\prime}
\ee
with $Q=\sqrt{\frac{q}{p}}$.
The diagonal boundary terms remain unchanged.

Although the Hamiltonian of the XYZ--chain with non--diagonal boundary terms
is known to be integrable \cite{Vega, Inami},
Bethe--Ansatz equations have not yet been obtained,
because it is not clear how to construct a reference state. 
Therefore, to study the effect of non--diagonal boundary terms,
we chose the XX--model with boundaries of the form given by eq.\ (\ref{Ham}),
because this model can be fermionized.
To be able to use the free fermion approach,
we introduce a new Hamiltonian $H_{long}$
which is bilinear in terms of fermionic creation and annihilation operators.
This approach has the major advantage that
we have complete control over the wave functions
for a large class of boundary parameters which enables us
to calculate correlation functions. 
Thus, we get a good handle on
a particularly interesting and important integrable model. 

As mentioned above, in order to treat the Hamiltonian given by 
eq. (\ref{Ham}) we transfer
the diagonalization problem to a new Hamiltonian which we obtain by appending 
one additional site at each end of the chain as in 
\cite{Bariev_Peschel}. This new Hamiltonian 
has the following expression
 \ba
\label{H_long}
H_{long}=\frac{1}{2}
\sum_{j=1}^{L-1} \Bigl[  \sp_j \sm_{j+1} + 
 \sm_j \sp_{j+1} \Bigr] &+&
\frac{1}{\sqrt{8}} \Bigl[
\a\,\sigma^x_0\sigma_1^- + \g\, 
\sigma^x_0\sp_1+\k\, \s^{z}_1\nn\\
&+&\b\,\sigma_L^+\sigma^x_{L+1} +  \d\, 
\sm_L \sigma^x_{L+1} +\rh \,\s^{z}_L \Bigr]
\ea
In this way, the boundary terms are also  bilinear
expressions in the $\sigma^+$ and $\sigma^-$ matrices. 
It is only after this transformation that we can
write and solve the problem in terms of free fermions. 
Since $\sigma_0^x$ and $\sigma_{L+1}^x$ commute
with $H_{long}$, the spectrum of 
$H_{long}$
decomposes into four sectors
$(++, +-, -+,--)$ corresponding to the eigenvalues $\pm 1$ of $\sigma_0^x$ and
$\sigma_{L+1}^x$. The original Hamiltonian corresponds to the (++)--sector.
A substantial part of this paper is devoted to showing how 
the eigenvalues of $H$ are obtained by projecting onto this sector.

The Hamiltonian $H_{long}$ which we introduced only as a means to treat 
the Hamiltonian $H$ is actually interesting in its own right 
as a quantum spin chain with boundary terms.       

In the field--theoretic approach the Hamiltonian $H_{long}$ is 
probably related  to
the decoupling point of the boundary sine--Gordon  model. The corresponding
boundary S--matrix has been calculated in \cite{Luca, Ameduri}. 

It is very likely that the Hamiltonian $H$ given by eq.\ (\ref{Ham})
can be applied to physical problems, since a simpler version of
this Hamiltonian has already found such applications.
Namely, the semi--infinite XX--chain with one $\sigma^x$--boundary term
mentioned before
was studied by Guinea \cite{Guinea} as a model for the dynamics 
of a particle in an external potential coupled to a dissipative 
environment. He also utilized free fermions 
and presented an explicit solution for the mobility of the particle
in the continuum limit. 
Afterwards this solution was used in
the study of transmission through resonant barriers 
and resonant tunnelling in an interacting one--dimensional electron gas,
cf.\  Kane and Fisher \cite{Kane_Fisher}. 
This type of system is studied in experiments with
quantum wires. The calculation is built on
a perturbative renormalization group analysis in different limits
(limits of a weak barrier and a strong barrier). 
By combining the results of these two limits the authors 
obtain the full phase diagram of the model. For one particular value
of the dimensionless conductance, they even
obtain an exact solution for the conductance through a
resonance by mapping the model onto the semi--infinite
XX--model with one $\sigma^{x}$--boundary term.

The starting point for our investigations of the XX--chain with
non--diagonal boundary terms is
the diagonalization of the Hamiltonian $H$
(eq.\ (\ref{Ham})). This 
problem is not only of mathematical interest, since the model 
has an interesting  physical content. Namely, 
as will be shown, boundary bound states appear and the non--trivial 
ground state expectation values of the $\sigma_j^x$--operators 
and the $\sigma_j^z$--operators exhibit a decay into the bulk which
can be predicted from conformal field theory. 
Furthermore,  the expressions for the partition functions formally coincide
with partition functions of a Coulomb gas with only magnetic 
charges or only electric charges, depending on the choice of
the boundary parameters in the Hamiltonian $H$. Additionally, 
the fermionic energies as well as the expressions for the ground state
energies show a logarithmic dependence on the lattice length 
for special choices of the boundary parameters.
The study of these 
physical properties is deferred to two subsequent articles.
Below we summarize the content of all articles and point out 
how the results of the present article enter into the further considerations.

In this first
article, we confine ourselves to studying the
integrable model
with non--diagonal boundary terms given by $H$
on a finite chain. This includes the
calculation of the spectrum and
the wave functions as well as the derivation of expressions for the one--
and two--point correlation functions for the $\sigma_j^x$--operators where
the subscript $j$ indicates the position on the chain. These results
are obtained in parallel for $H$ and $H_{long}$.
Let us briefly describe how we proceed:
We start by fermionizing the Hamiltonian $H_{long}$.
The spectrum of the original chain as well as the eigenvectors
can be retrieved from the spectrum  and the eigenvectors of 
$H_{long}$ by a projection technique which we derive in detail.
As a by--product, we solve
the eigenvalue problem for the quantum spin chain $H_{long}$. 
We demonstrate that after fermionization 
the problem of finding the eigenvalues  of the Hamiltonian $H_{long}$ 
is reduced to the problem of finding the zeros of a complex polynomial
of degree $2L+4$, which we write down explicitly. 
This polynomial, which might very well also appear in other contexts, 
has interesting algebraical properties. Namely,
for some choices of the parameters,
it can be factorized into cyclotomic polynomials.
We looked systematically for these factorizations 
since, apart from being of mathematical
interest,  these examples give access to an exact solution for the 
full spectrum
of the Hamiltonian, including exact expressions for the ground state energy.
(In the general, non--hermitian case,
we define the
ground state energy to be the one with 
the smallest real part.) Some of these examples are especially
interesting since the factorizations contain $L$--independent
factors which lead to $L$--independent eigenvalues of the Hamiltonian.
The corresponding eigenstates will be identified as boundary bound states 
in the next paper.
Furthermore, the ``cyclotomic'' examples furnish a 
reliable ansatz for
an approximative study of the zeros of the polynomial in the general case
which will be presented in the third paper.
As an additional result,  we get  exact 
formulas for the one-- and two--point correlation functions for
the $\sigma_j^x$--operators. The value of $\langle \sigma_{L+1}^x \rangle$
enters the projection mechanism mentioned above.

In the second article,  by using the results of the first paper,
we calculate one--point functions for the 
$\sigma_j^x$-- and the $\sigma_j^z$--operators for arbitrary position $j$
and lattice length $L$ for several of
the "exactly solvable" cases where the polynomial can be factorized
into cyclotomic polynomials. 
These one--point functions decay into the bulk with a power law 
typical of conformally invariant theories. Taking this point of view, 
we determine their critical exponents.

Furthermore, we make the connection between excitations 
with an $L$--independent
energy seen in this paper and boundary bound states.
This identification is made
on the one hand by studying the spatial profile of the special
fermionic excitations in comparison to
the spatial profile of other fermionic excitations, and on the other
hand by comparing them to boundary bound states found in the Bethe ansatz
for the XXZ--chain with diagonal boundary terms \cite{Skorik_Saleur}.
Boundary bound states originally have appeared in the
field--theoretic approach to the sine--Gordon model with boundary
interaction \cite{field_theory, Vega, Ameduri}.
In our case, it is surprising that they are related to 
special zeros of the complex polynomial as mentioned above and 
can therefore be found without invoking the field theory.

A further important new observation is related to the partition functions
in the thermodynamic limit. They will be presented in the third
article, where they will be derived by studying
approximative solutions of the polynomial equation for large values
of $L$ as mentioned above. The partition
functions correspond to conformally invariant systems, a behaviour which
we also found in our previous study of the 
totally asymmetric XX--chain with non--diagonal boundary terms
\cite{toy_paper}.
This observation is confirmed by the expansions of the exactly calculated 
ground state
energies for large $L$. From this expansion one can read off 
the conformal charge $c=1$ and obtain expressions for the surface free energy.
Moreover, the partition functions we find are the partition functions
of a Coulomb gas with only magnetic charges or only
electric charges. The phenomenon of finding only magnetic charges
is elucidated by
the construction of a pseudoscalar magnetic charge operator 
from the fermionic
number operators which commutes with the Hamiltonian for finite chains. 
Furthermore, for special choices of the boundary parameters, we find
a logarithmic $L$--dependence for the fermionic energies as well as
for the expression for the ground state energies. This may only happen 
if the Hamiltonian is non--hermitian.

\vspace{1cm}

The present article is very technical by nature. For those readers who are not
interested in all details of our calculations but would nevertheless like 
to use our results without reading the whole paper
we provide a guide in section 13, which does not, however, follow
the sections in a chronological way.
The others sections are
organized as follows: In section 2, we use 
the fermionization of the chain $H_{long}$
to reduce the eigenvalue problem of this Hamiltonian to the eigenvalue
problem of a matrix $M$ of dimension $(2L+4)\times(2L+4)$
whose eigenvalues correspond to the fermionic energies.
We derive some general properties of the eigenvectors of $M$ 
(which will be needed in the sections 10 and 11) before 
showing, in section 3, how the eigenvectors corresponding to the non--zero
eigenvalues of $M$ can be calculated explicitly.
The solution of the eigenvalue problem of $M$  leads to a complex polynomial
(which corresponds to the characteristic polynomial of $M$)
whose zeros determine all eigenvalues and eigenvectors of $M$. This
polynomial is presented in section 4. 
Section 5 is devoted
to the study of the factorization properties of this polynomial.
By constraining the total number of cyclotomic factors,
we systematically determined the boundary parameters for which
the polynomial factorizes into cyclotomic
polynomials. Some of these  cases 
are actually one--parameter families of solutions.  
In section 6, 
we show for two examples how the full spectrum 
of $H_{long}$ is obtained from the factorized form of the polynomial.
Section 7 contains the exact expressions for the ground state energies
of all examples where the polynomial factorizes into cyclotomic
polynomials. 
In section 8, we present one example of a Hamiltonian with
asymmetric bulk terms where it is also possible to 
calculate the full spectrum and the ground state energy exactly for arbitrary
values of $L$.
In section 9, we derive the projection mechanism which is needed
to obtain the spectrum of the original Hamiltonian from $H_{long}$.
To derive the projection mechanism we need
the value of the one--point function of the $\sigma_j^x$--operator
 at the point $j=L+1$.
We express the
one-- and two--point correlation functions of $\sigma_j^x$ in terms
of Pfaffians in section 10. 
In the cases where the Hamiltonian $H$  has no $\sigma^z$
boundary terms or fulfils the condition $\a =\g$ and $\b =\d$, 
we further reduce these Pfaffians to determinants of a certain matrix.
These  expressions  will  be needed for the 
calculation of spatial profiles in our second article. 
In section 11, we calculate the above mentioned 
value of the one--point function of 
$\sigma_{j}^x$ at the point $j=L+1$ in the cases where the 
Hamiltonian is: a) hermitian, b) has no $\sigma^z$ boundary terms or
c) fulfils the condition $\a =\g$ and $\b =\d$. Inputting this result we invoke 
the projection mechanism and 
present the ground state energies for the original Hamiltonian $H$
in the ``exactly solvable'' cases which
additionally satisfy at least one of the afore mentioned conditions a)--c)
 in section 12.
We conclude this article with a discussion of our results
in section 14. In an appendix we show how to find
the eigenvectors of the matrix $M$ corresponding to the eigenvalue zero.
We derive the conditions for the appearance of
zero modes in the spectrum of $H_{long}$ and determine respective restrictions
for the boundary parameters.

\section{Diagonalization of the Hamiltonian}
In this section, we present the general formalism we use for the 
diagonalization of the XX-model with boundary terms
defined by eq. (\ref{Ham}).  $H$ can be diagonalized in terms of 
free fermions if it can be written 
as a bilinear expression in $\sigma^{\pm}$-matrices, since 
standard fermionization 
techniques can then be applied \cite{Jordan_Wigner, LSM}.

To obtain a bilinear expression in $\sigma^{\pm}$-matrices
for $H$ we add one lattice site
at each end of the chain, site $0$ and site $L+1$ as in \cite{Bariev_Peschel}. 
Notice that the terms containing $\s^{z}$ 
do not have to be changed.
The Hamiltonian now reads
\ba
H_{long}=\frac{1}{2}
\sum_{j=1}^{L-1} \Bigl[  \sp_j \sm_{j+1} + 
 \sm_j \sp_{j+1} \Bigr] &+&
\frac{1}{\sqrt{8}} \Bigl[
\a\,\sigma^x_0\sigma_1^- + \g\, 
\sigma^x_0\sp_1+\k\, \s^{z}_1\nn\\
&+&\b\,\sigma_L^+\sigma^x_{L+1} +  \d\, 
\sm_L \sigma^x_{L+1} +\rh \,\s^{z}_L \Bigr]
\ea
As $\sigma_0^x$ and $\sigma_{L+1}^x$ commute
with $H_{long}$, the spectrum of 
$H_{long}$
decomposes into four sectors
$(++, +-, -+,--)$ corresponding to the eigenvalues $\pm 1$ of $\sigma_0^x$ and
$\sigma_{L+1}^x$.
Notice that the projection of $H_{long}$ onto a fixed sector 
$(\epsilon_1,\epsilon_2)$
has the same eigenvalues as the original Hamiltonian with the
choice of the parameters $\epsilon_1 \;\a$, $\epsilon_1 \;\g$, $\epsilon_2 \;\d$
and $\epsilon_2 \;\b$ so that by diagonalizing $H_{long}$ one 
simultanously treats 
four different Hamiltonians $H$.
The eigenvectors of the original choice of the parameters $\a, \g, \d$ and $\b$
can be retrieved
by projecting onto the $(++)$-sector as described
in section 9.

Furthermore notice that the $(++)$ sector and the $(--)$ sector respectively
the $(+-)$ sector and the $(-+)$ sector can be interchanged by using the 
following transformation which leaves $H_{long}$ invariant:
\be
\label{symm}
\sigma^x_j \rightarrow -\sigma^x_j \;\;\;\;\;\;\;\;\;\;\;\;
\sigma^y_j \rightarrow -\sigma^y_ j\;\;\;\;\;\;\;\;\; \;\;\; 
\sigma^z_j \rightarrow \sigma^z_j\;\;\;\;\;\;\;\;\; \;\;\;
j=0,\ldots,L+1 \quad
\ee
It maps any eigenvector $|\Psi \rangle$ of $H_{long}$ from the 
$(\epsilon_1,\epsilon_2)$--sector onto an 
eigenvector $|\Psi \rangle ^{\prime}$ of $H_{long}$ with the same eigenvalue 
lying in the sector $(-\epsilon_1,-\epsilon_2)$. Therefore each
eigenvalue of $H_{long}$ is at least twofold degenerate. In the
fermionic language, the above symmetry manifests itself as a zero mode.

In the next section, we will show that the diagonalization of 
$H_{long}$ can be reduced to finding the eigenvalues and the
eigenvectors of a  $(2L+4)\times(2L+4)$--matrix which will be
denoted by $M$.
After studying general properties of the eigenvectors, we
will describe in section 3 how they can be obtained in an explicit form. 
The eigenvectors and 
the eigenvalues of the matrix $M$ are determined by the zeros of a polynomial 
which will be given in section 4. 

\subsection{Diagonalization of $H_{long}$}
Adopting the Majorana representation of the lattice
$s=1/2$ spin operators as in \cite{Haye}, set 
\be
\label{tau}
\tau_j^{+,-} = (\prod_{i=0}^{j-1} \sigma^z_i)\sigma_j^{x,y}
\ee
These operators obey the anticommutation relations of a Clifford--algebra
$ \{ \tau_m^{\mu}, \tau_n^{\nu} \} = 2 \delta_{nm}^{\mu \nu}$.
Rewriting $H_{long}$ in terms of $\tau_j^{+,-}$, we obtain the following
bilinear expression 
\be
\label{hI}
\fl
H_{long}
=-\sum_{\mu,\nu=\pm 1}\sum_{j=1}^{L-1} F^{\mu,\nu}\tau^{\mu}_j
\tau^{\nu}_{j+1} + G^{\mu,\nu}\tau^{\mu}_0\tau^{\nu}_{1}+
K^{\mu,\nu}\tau^{\mu}_L\tau^{\nu}_{L+1} + I^{\mu,\nu}\tau^{\mu}_1\tau^{\nu}_{1}
+J^{\mu,\nu}\tau^{\mu}_L\tau^{\nu}_{L}
\ee
where
\ba
\label{matrices}
\fl
G & =&\frac{1}{2}
\left(
\begin{array}{ll}
        \frac{1}{\sqrt{8}}(\a - \g) & \frac{\i}{\sqrt{8}}(\a 
+ \g) \\
        0 & 0
\end{array}
 \right),\quad
K=\frac{1}{2}
\left(
\begin{array}{ll}
        0 & \frac{\i}{\sqrt{8}}(\b + \d) \\
        0 & \frac{1}{\sqrt{8}}(\b- \d)
\end{array}
 \right)
\nn \\
\fl
F &=& \frac{1}{4}
\left(
\begin{array}{ll}
        0 & \i \\
        -\i & 0
\end{array}\right),\quad
I = \frac{1}{2}
\left(
\begin{array}{ll}
        0 & -\frac{\i}{\sqrt{8}} \k \\
        \frac{\i}{\sqrt{8}} \k & 0
\end{array}
 \right),\quad
J=\frac{1}{2}
\left(
\begin{array}{ll}
        0 & -\frac{\i}{\sqrt{8}} \rh  \\
        \frac{\i}{\sqrt{8}} \rh & 0
\end{array}
 \right) 
\label{matrix}
\ea
Here we chose the basis such that the matrices above have the
general form $
A= \left(
\begin{array}{ll}
        A^{--} & A{-+} \\
        A^{+-} & A^{++}
\end{array}
 \right)$
where $A$ is one of the matrices in (\ref{matrices}).
Now we apply a linear transformation to the $\tau_j^{+,-}$ operators
to obtain another set $T_n^+, T_n^-$ 
of Clifford operators again satisfying
\be
\label{Clifford}
\{ T_m^{\mu}, T_n^{\nu} \} = 2 \delta_{nm}^{\mu \nu}
\ee
Let
\be
\label{lintrafo}
T_n^{\gamma}=\sum_{j=0}^{L+1}\sum_{\mu=\pm 1} (\psi^{\gamma}_n)^{\mu}_j
\tau^{\mu}_j
\ee
be the explicit form of this linear transformation  
where $\gamma = \pm 1$.  
One can choose this linear transformation in such a way that in
 terms of these new Clifford operators $H_{long}$ takes 
the simple form
\be
\label{hfer}
H_{long}=\sum_{n=0}^{L+1}\Lambda_n\, \i T^-_n T^+_n 
\ee
The commutation relations for the $T^-_n, T^+_n$ imply that the
operator $\i T^-_n T^+_n$ has eigenvalues $\pm 1$ so that
the spectrum of $H_{long}$ is given by all possible linear combinations
involving all $\Lambda_n$ with coefficients $+1$ or $-1$ and can be read off
eq.\ (\ref{hfer}).

Notice that the operators $T_n^-,T^{+}_n$ as defined by eq.\ (\ref{lintrafo})
are in general non--hermitian. However, according to a general
theorem for Clifford operators \cite{Clifford}, it is possible to apply a 
similarity transformation to the set of vectors $(\psi^{+}_n)$
and $(\psi^{-}_n)$ to obtain new hermitian Clifford operators 
$T^{- \prime}_n, T^{+ \prime}_n$ in terms of which the Hamiltonian 
also takes the form given by eq.\ (\ref{hfer}).
This will be discussed in detail
in the next section.

The coefficients $(\psi^{\gamma}_n)^{\mu}_j$
of eq.\ (\ref{lintrafo}) are constrained
by requiring that the  operators $T_n^{\gamma}$ obey
the anticommutation relations of eq.\ (\ref{Clifford}):
By computing
the commutator $[H_{long},T_n^{\pm}]$ using for
$H_{long}$ first the expression (\ref{hfer}), and then 
(\ref{hI}), and comparing both results,  one finds
that
the eigenvalues $\Lambda_n$ and the vectors
\be
\label{eigenvector}
\psi^{\gamma}_n=((\psi^{\gamma}_n)^-_0,(\psi^{\gamma}_n)^+_0,...,
(\psi^{\gamma}_n)^-_{L+1},
(\psi^{\gamma}_n)^+_{L+1}), \gamma=\pm
\ee
are given by the solutions of the following equations 
\be
\label{mpsi}
M \psi_n^+ = - \i \Lambda_n \psi_n^- \; , \;\;\;\;\;\;\;
M \psi_n^- =  \i \Lambda_n \psi_n^+
\ee
where $M$ is a $(2L+4)\times(2L+4)$ matrix given by
\be
\label{matrix_M}
M=
\left(
\begin{array}{cccccc}
0 &G\\
-G^T & 2 I & F\\
&-F^T&0&F\\
&&...&...&...\\
&&&-F^T&  2 J &K\\
&&&&-K^T&0\\
\end{array}
\right)\quad ,
\ee

Defining 
\be
\label{def_eigen}
\phi_n^{+} = \psi_n^+ - \i\psi_n^- \; , \;\;\;\;\;\; \phi_n^{-}=
\psi_n^+ + \i\psi_n^-
\ee
leads to the eigenvalue problem
\be
\label{eigenvalue_problem}
M\phi_n^{\pm} = \pm \Lambda_n \phi_n^{\pm}
\ee

Observe that $M$ has $2L+4$ eigenvalues although 
$H_{long}$ has only length $L+2$. This can be explained by 
considering eq. (\ref{eigenvalue_problem}).
As one can see, with the appearance of each eigenvalue $\Lambda_n$ 
we also get the negative eigenvalue $-\Lambda_n$. As mentioned above, 
the spectrum of $H_{long}$ is given by all linear combinations 
of $\Lambda_n$ with coefficients $\pm 1$ (see eq. (\ref{hfer}))
and thus can be retrieved 
from the eigenvalues of $M$ by choosing from each pair of eigenvalues
$\pm \Lambda_n$ one value as basis element for the ${\bf Z}_2$ -- linear
combinations. Later we will make this choice in a systematic way 
following a physical convention which consists of  choosing
as relevant energies the eigenvalues with positive real part. 

As can be seen directly from the form of $M$, at least two of the eigenvalues
$\Lambda_n$ are zero. The corresponding eigenvectors 
are given by $(0,1,0,0,\ldots,0)$ and $(0,0,\ldots,0,1,0)$. Since
the eigenvalues of $M$ occur in pairs $\pm \Lambda_n$, from which
only one value has to be taken, the  zero eigenvalues   
lead to one zero mode as  already mentioned above.
As we are going to see more explicitly, this zero mode does not appear as a
fermionic excitation in the
spectrum of $H$. This fact can be explained as follows:
Recall that in the case of $H_{long}$ the zero mode 
reflects the presence of the symmetry given by eq.\ (\ref{symm}) 
which interchanges
the  $(++)$ sector and the $(--)$ sector respectively
the $(+-)$ sector and the $(-+)$ sector. Since $H$ corresponds only to the
$(++)$ sector, it is clear that the above symmetry is not a symmetry
of $H$. Therefore the above zero mode does not appear in the spectrum of $H$.
In the following  we are going to call it spurious 
zero mode.

To express the spectrum of $H_{long}$ in terms of free
fermions, we will now write the
expression for the Hamiltonian in terms
of fermionic operators
$b_n$ and $a_n$ satisfying
\be
\label{glei_a}
\{b_n,a_{m} \} =
\delta_{n,m}\; ,\;\;\;\;\; \{ b_n, b_m \} = 0\; , \;\;\;\;
\; \{ a_{n}, a_m \} = 0
\ee
which are
obtained from the Clifford operators $T_n^{+}$ and $T_n^{-}$ by the following
transformation: 
\be
\label{ainT}
b_n = \half(T_n^+ + \i T_n^-); \;\;\;
a_n = \half(T_n^+ - \i T_n^-)
\ee
$H_{long}$ then reads
\be
\label{spectrum}
H_{long} = \sum_{n=0}^{L+1} 2\,
\Lambda_n\, b_n a_n -
\sum_{n=0}^{L+1}\Lambda_n
= \sum_{n=0}^{L+1} 2 \,\Lambda_n\, N_n + E_0
\ee
where $E_0$ is the ground state energy of the system and $N_n$ the
number operator (with eigenvalues $0$ and $1$)
for the fermion with energy $2 \Lambda_n$.

Notice that in the expression for the number operator $N_n$
in eq.\ (\ref{spectrum}) $b_n$ is 
equal to $a^{\dagger}_n$ if the operators $T^+_n$ and $T^-_n$
are hermitian.  As mentioned above, they always can be chosen to be
hermitian by applying a similarity transformation to the
vectors $(\psi^{+}_n)$
and $(\psi^{-}_n)$ in eq.\ (\ref{lintrafo}). 
At the same time, the operators $a_n$ and $b_n$ are then transformed
into new  operators
$a_n^{\prime}$ and $b_n^{\prime}$ which are adjoints of each other.

In eq.\ (\ref{spectrum}) we have defined the Fermi sea by summing
over all negative eigenvalues of $M$. 
Consequently, we have to choose the other half of the eigenvalues of $M$  
to form fermionic excitations above the Fermi sea. Here and in the
following we will use the convention that 
if a pair of eigenvalues has non--vanishing real part, we 
will denote the one with positive real
part by $\Lambda_n$. This choice leads 
to a ground state energy with the smallest real part.
In the case where the real part 
(but not the imaginary part) of $\Lambda_n$ is zero
one has the freedom of
choice to take either the eigenvalue with positive or the eigenvalue
with negative
imaginary part as a fermionic excitation above the
Fermi sea. This leads to an ambiguity in the value of the  imaginary
part of the ground state energy.
A similar problem occurs in the  calculations involving the eigenvectors
of the zero modes (e.\ g.\ in the calculation of one--point
functions of $\sigma$--operators). Namely, the zero eigenvalues of $M$
also occur in pairs ('$+0$' and '$-0$') and one can freely choose
which of these  two zero eigenvalues belongs to the Fermi sea
and which one corresponds to an excitation with zero energy.
In other words, one can choose which is the 
creation and which the annihilation operator corresponding to
the fermion with zero energy. 
We will come back to this point in \cite{paper2}.

The fermionic operators $a_m,b_m$ can be expressed in terms
of the $\tau_j^+,\tau_j^-$-operators by using the eigenvectors of $M$ and
eq.\ (\ref{lintrafo}) 
\be
\label{a}
a_m  = \half(T_m^+ - \i T_m^-) =
\half \sum_{j=0}^{L+1}\sum_{\mu=\pm 1} (\phi^{+}_m)^{\mu}_j
\tau^{\mu}_j
\ee
\be
\label{a_dagger}
b_m  =  \half(T_m^+ + \i T_m^-) =
\half \sum_{j=0}^{L+1}\sum_{\mu=\pm 1} (\phi^{-}_m)^{\mu}_j
\tau^{\mu}_j
\ee

Notice again that in general the expression given by eq. (\ref{a_dagger})
is not the adjoint of 
$a_m$ because the Hamiltonian is non-hermitian. However, if one transforms
the operators $T_n,T^+_n$ into hermitian operators as mentioned above,
the set of vectors $(\phi^{+}_m), (\phi^{-}_m)$ fulfils the 
conditions $(\phi^{+}_m)  = (\phi^{-}_m)^{*}$ and $b_m$ becomes the
adjoint of $a_m$.

\begin{subsection}{Orthogonality relations}

In the following, we make some general remarks on the given 
eigenvalue problem for the skew--symmetric
matrix $M=-M^t$ defined by eq.(\ref{eigenvalue_problem}).
We show that we can indeed find a linear transformation of the form 
given by eq.\ (\ref{lintrafo})
in terms of  
the vectors $(\psi^{\gamma}_n)^{\mu}_j$ (which are related to the 
eigenvectors of $M$ by eq.\ (\ref{def_eigen}))
such that  $a_k$ and $b_k$ of 
eq.(\ref{a_dagger}),(\ref{a})
satisfy eq.(\ref{glei_a}) or equivalently
that the $T_k^{\pm}$ of eq.(\ref{lintrafo}) 
are Clifford operators respectively, i.e. satisfy eq.(\ref{Clifford}),
which was assumed before deriving the eigenvalue equation.
The corresponding orthogonality relations for the eigenvectors of $M$
which are equivalent to the 
anti-commutation relations for the operators $a_k,b_k$
lead to further
relations between
 the eigenvectors (see eqs.(\ref{crucial}) and (\ref{crucial2}))
for special choices of the boundary parameters. 
They simplify the computation of correlation functions and are used for 
projecting to the $(++)$-sector of $H_{long}$. This will be the subject 
of the sections 9,10 and 11.

Let us first look at the case where the Hamiltonian is hermitian, i.e. $\alpha_-=\alpha_+^*$, 
$\beta_-=\beta_+^*$ and $\alpha_z, \beta_z \in \BRT$. This implies that $M$ has only purely imaginary
entries and is also hermitian. So its eigenvectors can be chosen
to form an orthogonal basis with respect to the standard scalar product.
Because $M^*=-M$ we have
\be
\phi_k^-\propto {\phi_k^+}^*
\label{bdagger=a}
\ee
 which can be directly seen by taking the complex 
conjugate of the equation $M\phi_k^+=\Lambda_k \phi_k^+$. 
Thus after an appropriate normalization of 
the eigenvectors the orthogonality condition for
the eigenbasis is equivalent to 
the relations which are necessary and sufficient to
define a set of fermionic operators (eq.(\ref{glei_a})):
\be
\sum_{j=0}^{L+1}\sum_{\gamma} (\phi^+_l)^{\gamma}_j(\phi^-_k)^{\gamma}_{j} = 2\delta_{lk} \quad ,
\label{ev2}
\ee
\be
\sum_{j=0}^{L+1}\sum_{\gamma} (\phi^+_l)^{\gamma}_j(\phi^+_k)^{\gamma}_{j}=
\sum_{j=0}^{L+1}\sum_{\gamma} (\phi^-_l)^{\gamma}_j(\phi^-_k)^{\gamma}_{j}=0 \quad .	
\label{ev0}
\ee
Note that due to eq.(\ref{bdagger=a}) $\phi_k^-$ and $\phi_k^+$ can always be
normalized so that $b_k$ and $a_k$ are mutually adjoint.
For any set of constants $c_k\in\BCT,c_k\neq 0$
the vectors $\psi_k^+=\frac{1}{2}(c_k\phi_k^+ + c_k^{-1}\phi_k^-)$ and
$\psi_k^-=\frac{\i}{2}(c_k\phi_k^+ - c_k^{-1}\phi_k^-)$ 
satisfy eqs.(\ref{mpsi}) and the orthogonality relations
\be 
\sum_{j=0}^{L+1}\sum_{\gamma} (\psi^{\mu}_l)^{\gamma}_j(\psi^{\nu}_k)^{\gamma}_j= \delta^{\mu\nu}_{lk}
\label{clifforth}
\ee 
and thus the $T^{\pm}$ defined in terms of the  $(\psi^{\gamma}_n)^{\mu}_j$
by  eq.(\ref{lintrafo}) are Clifford operators. 
If we define  $\bf \Psi$ to be the $(2L+4)\times (2L+4)$
 matrix consisting of the $2L+4$ vectors $\psi_k^{\pm}$,
we may rewrite eq.(\ref{clifforth}) as $\bf \Psi^t \Psi = 1$. This simply
reflects the fact that the 
automorphism group of the Clifford algebra is the orthogonal group. Note 
that here $\bf \Psi$ is 
not necessary real. However, due to eq.(\ref{bdagger=a})
 $\bf \Psi$ can always be made real by tuning the parameters $c_k$.
Since $\bf \Psi^t \Psi = 1$ implies $\bf \Psi \Psi^t = 1 $, we also obtain 
\be
\sum_{k=0}^{L+1}\sum_{\mu} (\psi^{\mu}_k)^{\gamma}_j(\psi^{\mu}_k)^{\nu}_i = \delta^{\gamma\nu}_{ij}
\label{kt}
\ee
or in terms of the components of the eigenvectors of $M$
\be
\sum_{k=0}^{L+1}\sum_{\mu} (\phi^{\mu}_k)^{\gamma}_j (\phi^{-\mu}_k)^{\nu}_i = 2\delta^{\gamma\nu}_{ij} \quad .
\label{kev}
\ee
Using these equations it is possible to invert 
eq.(\ref{lintrafo}) and eqs.(\ref{a_dagger}),(\ref{a}) respectively.
This is necessary to express the spin operators $\sigma^x,\sigma^y,\sigma^z$ 
in terms of ladder operators, which is needed for the calculation of
correlation functions and of the projection mechanism. We will use this form of the    
orthogonality relations in the next subsection, in order to derive some further relations
between the eigenvectors.

If the Hamiltonian is not necessarily
hermitian but all of the eigenvalues of $M$ are non--degenerate except for the 
 eigenvalue 0 corresponding to the eigenvectors $(0,1,0,0,\cdots)$ and 
$(0,0,\cdots,1,0)$, one can still show that eqs.(\ref{ev2}) and (\ref{ev0}) remain valid. 
In general the argument breaks down because $M$ is not necessarily
diagonalizable. This will become apparent in the sections 3--5.

Choosing the linear combinations 
\be
\phi_{0}^+ = (0,1,0,\cdots,0,\i,0) \quad , \quad 
\phi_{0}^- = (0,1,0,\cdots,0,-\i,0)
\label{spuriousev}
\ee
 as the eigenvectors corresponding 
to the eigenvalue $0$ we ensure that they also satisfy
the eqs.(\ref{ev2}) and (\ref{ev0}).
We now check  eqs.(\ref{ev2}) and (\ref{ev0}) for the other eigenvectors 
of $M$: 

First of all, let $\phi^+$ be a right eigenvector corresponding to the eigenvalue $\Lambda$, i.e. 
\be
M\phi^+ =\Lambda \phi^+
\label{rightev}
\ee
Because of $M=-M^t$ this eigenvector is also a left 
eigenvector corresponding to the eigenvalue $-\Lambda$, i.e.
\be
{\phi^+}^t M = -\Lambda {\phi^+}^t \quad ,
\label{leftev}
\ee 
This implies the existence of a right eigenvector $\phi^-$
 corresponding to the eigenvalue $-\Lambda$.

Now let $\phi_k$ and $\phi_l$ be eigenvectors corresponding 
to eigenvalues $\lambda_k$ and $\lambda_l$ where we do not restrict the
real parts of $\lambda_k$ and $\lambda_l$
to be positive or negative.
Using eqs.(\ref{rightev}) and (\ref{leftev}) we get 
\be
{\phi_k}^t \phi_l = \frac{-\lambda_k}{\lambda_l}{\phi_k}^t \phi_l \quad .
\label{eukprod}
\ee
So all products of the form ${\phi_k}^t \phi_l$ are zero if 
$\lambda_k\neq-\lambda_l$. This gives eq.(\ref{ev0}).

To proof the validity of eq.(\ref{ev2}) we additionally have to show that
in the case $-\lambda_k = \lambda_l$ the product ${\phi_k}^t \phi_l$ cannot vanish.
This can be done by considering ${\phi_l}^{\dagger} \phi_l$ which is always
different from zero if $\phi_l\neq 0$.
Now due to the assumption of non degenerate eigenvalues
 the eigenvectors form a basis and thus
${\phi_l}^*$ can be expressed in terms of eigenvectors ${\phi_j}$.
Using eq.(\ref{eukprod}) with $-\lambda_k = \lambda_l$ we have
\be
0 \neq \phi_l^{\dagger}\phi_l = {\phi_l^*}^t\phi_l = \sum_j a_j \phi_j^t \phi_l = a_k\phi_k^t\phi_l 
\quad.
\ee
The only term left in the expansion of the product 
${\phi_l}^{\dagger} \phi_l$ is  
proportional to the product ${\phi_k}^t \phi_l$ due to eq.(\ref{eukprod})
and therefore cannot
vanish.
So we can normalize the eigenvectors appropriately in order to satisfy
equation (\ref{ev2}). If there are degeneracies in the spectrum of
$M$,  the above proof can be generalized by using the biorthogonality of left
and right eigenvectors.

We want to point out that the ladder operators $a_k$ and $b_k$  
are not the adjoints of each other
 in general because the relation (\ref{bdagger=a}) is not valid 
in general. However, as already mentioned in section 2.1
 it is always possible to perform a similarity transformation in order
to achieve $b_k^{\dagger}=a_k$. This can be seen by choosing
 an arbitrary real symmetric and orthogonal
matrix ${\bf \Psi'}$. The transformed vectors
\be
\psi'^{\mu}_k={\bf \Psi' \Psi^t} \psi_k^{\mu} 
\label{clifftrafo}
\ee 
define a new set of Clifford operators $T_k'^{\pm}$ which are now hermitian. Hence the
operators $b'_k=\frac{1}{2}(T_k'^+ + \i T_k'^-)$ and $a'_k=\frac{1}{2}(T_k'^+ - \i T_k'^-)$
form a set of fermionic ladder operators satisfying $b_k'^{\dagger}=a'_k$.
Since the vectors $\phi_k'^{\pm}=\psi_k'^+ \i\mp \psi_k'^-$ are no longer eigenvectors 
of $M$ but of the transformed matrix 
\be
M'={\bf \Psi' \Psi^t}M{\bf \Psi \Psi'^t}
\ee
the transformation (\ref{clifftrafo}) corresponds to a similarity transformation of the
Hamiltonian $H_{long}$. 
 
\end{subsection}

\begin{subsection}{Special properties of eigenvectors}
In some cases 
there are further relations between the eigenvectors in addition
to the ones of eqs.(\ref{ev2}) and (\ref{ev0}). 
They are used
in the calculation of correlation functions and are even
necessary for the projection
method.
First notice that if $\alpha_z=\beta_z=0$, the 
matrices $M$ and $M^2$ respectively take the form
\be
M=\left(
\begin{array}{ccccc}
0&*&0&*&\cdots\\   
\,\!*&0&*&0&\\
0&*&0&*&\\
{*}&0&*&0&\\
\vdots&&&&\ddots\\
\end{array}
\right) \quad , \quad
M^2=\left(
\begin{array}{ccccc}
\,\!*&0&*&0&\cdots\\
0&*&0&*&\\
\,\!*&0&*&0&\\
0&*&0&*&\\
\vdots&&&&\ddots\\
\end{array}
\right)
\label{form}
\quad ,
\ee
where $*$ and $0$ both denote 2x2 matrices.
Note that $*$ is the notation for an arbitrary 2x2 matrix and is
not necessarily different from zero.
Looking at eq.\ (\ref{form}) we see that we can choose an 
eigenvector $\psit_k^+$ 
 of $M^2$ with eigenvalue $\Lambda_k^2$, i.e.
\be
M^2 \psit_k^+ = \Lambda_k^2 \psit_k^+  \label{evm2}
\ee
 which satisfies
\be
(\psit_k^+)^{\pm}_i= 0  \quad \mbox{for $\i$ odd} \quad . 
\label{psip}
\ee
Now we define $\psit_k^-$ by
\be
M\psit_k^+= -\i \Lambda_k \psit_k^- \label{psim}
\ee
which is also an eigenvector of $M^2$ with eigenvalue $\Lambda_k^2$.
Note that this definition does not work if $\Lambda_k = 0$ and 
thus we have to exclude in the following $k=0$ which 
labels the eigenvectors corresponding
to the spurious zero mode.
Using eqs.(\ref{form}) and (\ref{psip}) we obtain 
\be
(\psit_k^-)^{\pm}_i = 0 \quad \mbox{for i even} \quad .
\ee

Due to eqs.(\ref{evm2}) and (\ref{psim}) 
$\psit_k^+$ and $\psit_k^-$ also satisfy eq.(\ref{mpsi})
and thus we obtain eigenvectors 
 $\phit_k^{\pm}=\psit_k^+\mp \i \psit_k^-$ of $M$ satisfying 
\be
(\phit^+_k)^{\mu}_j = (-1)^j (\phit^-_k)^{\mu}_j \quad .
\label{nodiag}
\ee
Therefore
for each pair of vectors $\phi_k^+,\phi_k^-$ satisfying eq.(\ref{ev2}),
 there exists 
a constant $c\in \BCT$ such that 
\be
(\phi^+_k)^{\mu}_j = c (-1)^j (\phi^-_k)^{\mu}_j \quad .
\label{nodiagn}
\ee
Now eq.(\ref{kev}) can be rewritten as 
\be
(1+(-1)^{i+j})
\sum_{k=0}^{L+1} (\phi^+_k)^{\gamma}_j (\phi^-_k)^{\nu}_i = 2\delta^{\gamma\nu}_{ij} 
\label{cru}
\ee
and we end up with
\be
\sum_{k=0}^{L+1} (\phi^+_k)^{\gamma}_j (\phi^-_k)^{\nu}_i = \delta^{\gamma\nu}_{ij} \quad , \quad
\mbox{for} \quad i+j \quad \mbox{even} .
\label{crucial}
\ee
Since eq.(\ref{nodiagn}) is not valid for $k=0$ and odd $L$ (see eq.(\ref{spuriousev}))
we exclude $i=0,\nu=+,j=L+1,\gamma=-$ and $j=0,\gamma=+,i=L+1,\nu=-$
 in the eqs.(\ref{cru}) and (\ref{crucial}).

If diagonal boundary terms are included, and if 
$\alpha_-=\alpha_+$ and $\beta_-=\beta_+$,
the eigenvectors again have a special property. 
In this case $M$ and $M^2$ have also the form of eq.(\ref{form}), but
now $*$ and $0$ just denote complex numbers and,  in place of
eq.\ (\ref{nodiagn}), we obtain 
\be
(\phi^+_k)^{\pm}_j = \pm c' (\phi^-_k)^{\pm}_j \quad ,
\label{nosigmayn}
\ee
which gives 
\be
\sum_{k=0}^{L+1} (\phi^-_k)^{\mu}_i (\phi^+_k)^{\mu}_j = \delta_{ij} \quad .
\label{crucial2}
\ee
Both of the equations (\ref{crucial}) and (\ref{crucial2})
 will be used in the sections 10 and 11.

Note that the proof of eqs.(\ref{crucial}) and (\ref{crucial2}) shown above is
not valid if there are degeneracies or zero modes on top of the spurious zero mode
 in the spectrum of $M$. 
However, one can show that it is always possible 
to build appropriate linear combinations of the eigenvectors
corrsponding to the same eigenvalue such that
eqs.(\ref{crucial}) and (\ref{crucial2}) remain valid in
 addition to eqs.(\ref{ev2}) and (\ref{ev0}).
This is not automatically true and
therefore in explicit calculations one should take care in choosing
the right linear combination of eigenvectors corresponding
to the degenerate eigenvalues.

If both conditions $\alpha_-=\alpha_+, \beta_-=\beta_+$ and 
$\alpha_z=\beta_z=0$ are satisfied at the same time,
then both of the equations (\ref{crucial}) and (\ref{crucial2}) 
can be satisfied simultaneously.
By comparing eqs.(\ref{nodiagn}) and (\ref{nosigmayn})
we obtain the following relation
\be
(\phi_k^-)_j^{\pm}=\left\{
\begin{array}{cc}
\pm \frac{c'}{c}  (\phi_k^-)_j^{\pm} & \mbox{for $j$ even} \\
\mp \frac{c'}{c}  (\phi_k^-)_j^{\pm} & \mbox{for $j$ odd}
\end{array}
\right.
\quad .
\ee
Thus $\frac{c'}{c}$ is either $1$ or $-1$ because otherwise the vectors
$\phi^{\pm}_k$ would vanish.

Let us briefly summarize what we have obtained so far. If $H$ is hermitian,
the fact that $M = -M^t$ leads to
eqs.(\ref{ev2}),(\ref{ev0}) which are equivalent to
the anticommutation relations (\ref{glei_a}) of the ladder operators given by
 eqs.(\ref{a_dagger}), (\ref{a}). 
These equations are still valid if $H$ is
non-hermitian, but diagonalizable.
Some additional properties of the eigenvectors have been derived for
special choices of the boundary parameters
(eq. (\ref{crucial}) for $\alpha_z = \beta_z = 0$ and eq. (\ref{crucial2}) 
for $\alpha_- = \alpha_+, \beta_- = \beta_+$) which will be
used in the sections 10 and 11.
\end{subsection}

\section{Calculation of the eigenvectors of the matrix $M$}

In the previous section we have shown some general properties of 
the eigenvectors without
computing them explicitly. This computation is the subject of this section.
We will also show how to obtain the characteristic polynomial
 which gives the eigenvalues of $M$.
This polynomial will be treated extensively in the following sections.
Since $M$ is non hermitian in general we will also discuss the 
diagonalizability of $M$.

The eigenvalue problem given by eq.(\ref{eigenvalue_problem})
is equivalent to a set of recurrence relations. 
Using the notation given by eqs.(\ref{def_eigen}) and (\ref{eigenvector})
for the eigenvectors $(\phi_k^{\pm})$ of $M$ let us
first look at the bulk part :
\be
\begin{array}{rr}
\frac{\i}{4}((\phi_k^{\pm})_j^+ + (\phi_k^{\pm})_{j+2}^+)=\pm\Lambda_k (\phi_k^{\pm})_{j+1}^-  \\
-\frac{\i}{4}((\phi_k^{\pm})_j^-+(\phi_k^{\pm})_{j+2}^-)=\pm\Lambda_k (\phi_k^{\pm})_{j+1}^+         
&
\end{array}
\quad (1\leq j \leq L-2) \quad .
\label{bulkeq}
\ee  
These bulk equations (\ref{bulkeq}) can be decoupled by
defining 
\be
\vp_j=(\phi_k^{\pm})_j^- + \i (\phi_k^{\pm})_j^+ \quad , \quad
\vpd_j=(\phi_k^{\pm})_j^- - \i (\phi_k^{\pm})_j^+
\label{varphi}
\ee
which gives
\be
\begin{array}{rr}
\frac{1}{4}(\vp_j + \vp_{j+2})=\lambda \vp_{j+1}  \\
-\frac{1}{4}(\vpd_j+\vpd_{j+2})=\lambda \vpd_{j+1}
&\quad ,
\end{array}
\label{bulkeqd}
\ee
Here $\lambda=\pm\Lambda_k$ and the functions  
$\vp_j$ and $\vpd_j$ refer to
$(\phi_k^{+})_j^-$ and $(\phi_k^{+})_j^+$ for $\lambda=\Lambda_k$
and to $(\phi_k^{-})_j^-$ and $(\phi_k^{-})_j^+$ for $\lambda=-\Lambda_k$.
{}From now on we will keep $k$ fixed and omit all subscripts
referring to $k$.\\
Next we treat the left boundary, one obtains
\ba
\begin{array}{l}
\vp_0=\vpd_0 \\
\lambda\vp_0=\frac{1}{\sqrt{32}}(\alpha_-\vp_1 - \alpha_+ \vpd_1) 
\end{array} \label{leftb1}\\
\begin{array}{l}
\lambda\vp_1=  \frac{1}{\sqrt{8}}(\alpha_+ \vp_0 -\alpha_z\vp_1) + \frac{1}{4}\vp_2 \\
\lambda\vpd_1= \frac{1}{\sqrt{8}}(\alpha_z \vpd_1 - \alpha_-\vpd_0) - \frac{1}{4}\vpd_2 
\quad . \label{leftb2}\\
\end{array}
\ea
{From} the right boundary one gets
\ba
\begin{array}{l}
\lambda \vp_L = \frac{1}{\sqrt{8}}(\beta_+\vp_{L+1}-\beta_z \vp_L) + \frac{1}{4}\vp_{L-1} \\
\lambda \vpd_L = \frac{1}{\sqrt{8}}(\beta_z\vpd_L-\beta_-\vpd_{L+1}) -\frac{1}{4}\vpd_{L-1}\\
\end{array} \label{rightb1}\\
\begin{array}{l}
\lambda \vp_{L+1}=\frac{1}{\sqrt{32}}(\beta_+\vpd_L +\beta_-\vp_L) \\
\vp_{L+1}=-\vpd_{L+1} \quad . \label{rightb2}\\
\end{array}
\ea
Note that we have excluded explicitly
the  eigenvectors  $(0,1,0,...)$ and $(...,0,1,0)$, which always exist,
from
the set of solutions of the boundary equations above 
by setting $\vp_0=\vpd_0$ and $\vpd_{L+1}=-\vpd_{L+1}$. Thus we will 
obtain at most $2L+2$ linearly independent solutions instead of $2L+4$.

The general solution of the bulk equations (\ref{bulkeqd}) for $\lambda\neq\pm\frac{1}{2}$ is given by
\be
\label{solution_bulk}
\vp_j=ax^j+bx^{-j}\qquad , \qquad \vpd_j=g(-x)^j + f(-x)^{-j} \quad, \label{bulks}
\ee
where $1\leq j \leq L$ and up to now $a,b,g,f$ are free parameters which are independent of $j$.
The new variable $x$ 
is related to the eigenvalue $\lambda$ via 
\be
\label{x_lambda}
\lambda=\frac{1}{4}(x+x^{-1}) \quad .
\ee
For $\lambda=\pm\frac{1}{2}$ the general solution is
\be
\vp_j=a(\pm 1)^j + b(\pm 1)^j j \quad ,\quad  \vpd_j=g(\mp 1)^j + f(\mp 1)^j j \quad .
\label{sbulks}
\ee 
The four parameters $a,b,g,f$, the undetermined components $\vp_0,\vpd_0,\vp_{L+1},\vpd_{L+1}$
 and the eigenvalues $\lambda$
are all fixed by the boundary equations (\ref{leftb1}-\ref{rightb2}) ---
 up to the normalization constants of the eigenvectors.
Namely, plugging eq.(\ref{solution_bulk}) into eqs.(\ref{leftb1})-(\ref{rightb2}) we obtain a
homogeneous system of eight linear equations with the unknowns $a,b,g,f$ and
 $\vp_0,\vpd_0,\vp_{L+1},\vpd_{L+1}$. The condition for the existence of nontrivial
solutions of this system is given by the vanishing of the
determinant of the corresponding $8\times 8$ matrix. This defines 
a polynomial equation in the variable $x$ which yields all 
eigenvalues $\lambda$.
Note that for $x=\pm 1$
the $8\times 8$ system of equations has always the non--trivial solution
 $a=-b,\ g=-f$. 
This corresponds to the zero vector $\vp_j=\vpd_j=0\ \forall j$.
To compensate this fact we divide 
the polynomial by $(1-x^2)^2$. The treatment of the resulting
polynomial equation will be the subject
of the next section (see eq.(\ref{pol})).  

In the following we will show how to obtain the eigenvectors for
$\lambda\neq\pm\frac{1}{2}$ and $\lambda\neq 0$ which may be viewed as an alternative
way to obtain the secular equation. 
Substituting  (\ref{leftb1}) into (\ref{leftb2}) using the identity
$a=\vp_1 x^{-1} - b x^{-2}$ and 
$g=-\vpd_1 x^{-1} - f x^{-2}$ (see eq.(\ref{solution_bulk}))
renders $b$ and $f$ as functions of $\vp_1$ and $\vpd_1$, i.e.
\be
b=\frac{1}{1-x^{-2}} \left[
 \left(\frac{\alpha_-\alpha_+}{x+x^{-1}}-\sqrt{2}\alpha_z-x^{-1}\right)\vp_1
-\frac{\alpha_+^2}{x+x^{-1}}\vpd_1 
\right]
\label{b}
\ee
\be
f=\frac{1}{x^{-2}-1} \left[ 
 \left(\frac{\alpha_-\alpha_+}{x+x^{-1}}+\sqrt{2}\alpha_z-x^{-1}\right)\vpd_1
-\frac{\alpha_-^2}{x+x^{-1}}\vp_1
\right]
\ee
Thus $a$ and $g$ are given by
\be
a=\frac{1}{1-x^2} \left[
\left(\frac{\alpha_-\alpha_+}{x+x^{-1}}-\sqrt{2}\alpha_z-x\right)\vp_1
-\frac{\alpha_+^2}{x+x^{-1}}\vpd_1
\right] 
\ee
\be
g=\frac{1}{x^2-1} \left[
\left(\frac{\alpha_-\alpha_+}{x+x^{-1}}+\sqrt{2}\alpha_z-x\right)\vpd_1
-\frac{\alpha_-^2}{x+x^{-1}}\vp_1
\right]\label{g}
\ee
{}From the right boundary, by substituting  (\ref{rightb2}) 
into (\ref{rightb1}), we see that furthermore
\be
\label{right_boundary}
\vp_{L-1} + \left(\frac{\beta_-\beta_+}{4\lambda}-\sqrt{2}\beta_z-4\lambda\right)\vp_L
+ \frac{\beta_+^2}{4\lambda} \vpd_L = 0
\ee
\be
\label{right_boundary2}
\vpd_{L-1} - \left(\frac{\beta_-\beta_+}{4\lambda}+\sqrt{2}\beta_z-4\lambda\right)\vpd_L
- \frac{\beta_-^2}{4\lambda} \vp_L = 0 
\ee
Using eqs.(\ref{b}-\ref{g}) and eq.(\ref{bulks}) in the equations (\ref{right_boundary})
and (\ref{right_boundary2}) we get a linear system of equations of the form
\be
\label{m}
\left(
\begin{array}{ll}
\Omega_{11} & \Omega_{12}  \\
\Omega_{21} & \Omega_{22}
\end{array}
\right)
\left(
\begin{array}{l}
\vp_1 \\
\vpd_1 
\end{array}
\right) = 0
\ee
where the $\Omega_{ij}$ are the following 
functions of $x$ and the 6 boundary parameters 
$\alpha_{\pm},\beta_{\pm},\alpha_z$ and $\beta_z$
(Note that $\lambda$ is a function of $x$ according to eq.(\ref{x_lambda})):
\begin{eqnarray}
\fl
\Omega_{11} = & 
\frac{x^{-L}}{1-x^{-2}} & \left[
\left(\frac{\beta_-\beta_+}{4\lambda}-\sqrt{2}\beta_z-x^{-1}\right)
\left(\frac{\alpha_-\alpha_+}{4\lambda}-\sqrt{2}\alpha_z-x^{-1}\right)
+ (-1)^L\frac{(\beta_+\alpha_-)^2}{4\lambda^2}
\right] + \nonumber \\
\fl
& 
\frac{x^L}{1-x^2} &\left[
\left(\frac{\beta_-\beta_+}{4\lambda}-\sqrt{2}\beta_z-x\right)
\left(\frac{\alpha_-\alpha_+}{4\lambda}-\sqrt{2}\alpha_z-x\right)
+ (-1)^L\frac{(\beta_+\alpha_-)^2}{16\lambda^2}
\right]
\label{omega11}
\end{eqnarray}
\begin{eqnarray}
\fl
\Omega_{12} = &
\frac{x^{-L}}{x^{-2}-1} & \left[
\left(\frac{\beta_-\beta_+}{4\lambda}-\sqrt{2}\beta_z-x^{-1}\right)
\frac{\alpha_+^2}{4\lambda} 
+(-1)^L \left(\frac{\alpha_-\alpha_+}{4\lambda}+\sqrt{2}\alpha_z-x^{-1}\right)
\frac{\beta_+^2}{4\lambda}
\right] + \nonumber \\
\fl
&
\frac{x^L}{x^2-1} & \left[
\left(\frac{\beta_-\beta_+}{4\lambda}-\sqrt{2}\beta_z-x\right)
\frac{\alpha_+^2}{4\lambda}
+(-1)^L \left(\frac{\alpha_-\alpha_+}{4\lambda}+\sqrt{2}\alpha_z-x\right)
\frac{\beta_+^2}{4\lambda}
\right]  
\label{omega12}
\end{eqnarray}
\begin{eqnarray}
\fl
\Omega_{21} = &
\frac{x^{-L}}{1-x^{-2}} & \left[
\left(\frac{\alpha_-\alpha_+}{4\lambda}-\sqrt{2}\alpha_z-x^{-1}\right)
\frac{\beta_-^2}{4\lambda}
+(-1)^L \left(\frac{\beta_-\beta_+}{4\lambda}+\sqrt{2}\beta_z-x^{-1}\right)
\frac{\alpha_-^2}{4\lambda}
\right] + \nonumber \\
\fl
&
\frac{x^L}{1-x^2} & \left[
\left(\frac{\alpha_-\alpha_+}{4\lambda}-\sqrt{2}\alpha_z-x\right)
\frac{\beta_-^2}{4\lambda}
+(-1)^L \left(\frac{\beta_-\beta_+}{4\lambda}+\sqrt{2}\beta_z-x\right)
\frac{\alpha_-^2}{4\lambda}
\right]
\label{omega21}
\end{eqnarray}
\begin{eqnarray}
\fl
\Omega_{22} = &
\frac{x^{-L}}{x^{-2}-1} & \left[
\frac{(\alpha_+\beta_-)^2}{16\lambda^2}+ 
(-1)^L\left(\frac{\beta_-\beta_+}{4\lambda}+\sqrt{2}\beta_z-x^{-1}\right)
\left(\frac{\alpha_-\alpha_+}{4\lambda}+\sqrt{2}\alpha_z-x^{-1}\right)
\right] + \nonumber \\
\fl
&
\frac{x^L}{x^2-1} & \left[
\frac{(\alpha_+\beta_-)^2}{16\lambda^2}+
(-1)^L\left(\frac{\beta_-\beta_+}{4\lambda}+\sqrt{2}\beta_z-x\right)
\left(\frac{\alpha_-\alpha_+}{4\lambda}+\sqrt{2}\alpha_z-x\right)
\right]
\label{omega22}
\end{eqnarray}
The necessary condition to have non-trivial solutions 
is 
\be
\Omega_{11}\Omega_{22}-\Omega_{12}\Omega_{21}=0 \quad .
\label{detomega} 
\ee
This condition is equivalent to the polynomial equation which is obtained  
from the homogeneous $8 \times 8$ system of linear equations mentioned above.

The construction of eigenvectors shown here is not valid for $\lambda=\pm\frac{1}{2}$ and
$\lambda=0$.
But one can show that the eigenvectors for $\lambda=\pm\frac{1}{2}$ can be obtained by
\be
\label{lim}
\vp_j=\lim_{x\to\pm 1} (a x^j + b^{-j}) \quad ,
 \quad \vpd_j=\lim_{x\to\pm 1}( g(-x)^j + f(-x)^{-j}) \quad ,
\ee
where $a,b,g,f$ are given by eqs.(\ref{b})-(\ref{g}).
Using de L'Hospital's rule one recovers the form of eq.(\ref{sbulks}).
 The vector components
$\vp_1$ and $\vpd_1$ are again given as solutions of the $2 \times 2$ system (\ref{m})
using $x=\pm 1$.

The solution of the $2 \times 2$
 linear system (\ref{m}) is straightforward for a given
set of boundary parameters and
a given value of $x$. It cannot be given in a unique form
 because some of the $\Omega_{ij}$ might vanish. 
We will give the explicit form of the eigenvectors for some special choices of boundary 
parameters in \cite{paper2} where we are going to calculate the expectation values
of $\sigma_j^z$ and $\sigma_j^x$ where $j$ denotes the position on the lattice.
 
If all $\Omega_{ij}$ vanish the corresponding
eigenvalue is at least twofold degenerate and we obtain two linearly
 independent eigenvectors
since $\vp_1$ and $\vpd_1$ can be chosen independently of each other.
On the other hand, if a zero of the polynomial is twofold degenerate,
it is not clear that all $\Omega_{ij}$ vanish. 
This comes from the fact that the 
Hamiltonian is in general non-hermitian and might be non--diagonalizable.

We would like to point out that the appearance of
an eigenvalue $\lambda\neq 0$ which is more than twofold 
degenerate would prove that $M$ is non--diagonalizable.
This is indeed the case for some special choices of the boundaries for a given lattice length $L$. 
This can be seen by looking at the factorizations of the polynomial obtained  
in section 5. 

Up to now we have shown how to construct the eigenvectors
of the matrix M defined by eq.(\ref{matrix_M}). For
 $\lambda\neq 0$ the components of the
eigenvectors are given by
\be
(\phi_k^{\pm})_j^- = \frac{1}{2} (\vp_j + \vpd_j) \quad , \quad
(\phi_k^{\pm})_j^+ = - \frac{\i}{2} (\vp_j - \vpd_j)
\quad ,
\label{varphi2}
\ee
where $\vp_j$ and $\vpd_j$ have the form of eq.(\ref{solution_bulk}) for
$\lambda\neq \pm \frac{1}{2}$ and  are given by eq.(\ref{lim}) for
$\lambda = \pm \frac{1}{2}$. 
The parameters $a,b,g$ and $f$ are given by eqs.(\ref{b}-\ref{g}).
Finally $\lambda$ has to be determined from eq. (\ref{detomega})
as a function of $x$ (see eq. (\ref{x_lambda})).
The vanishing of the determinant in eq.(\ref{detomega})
leads to a complex polynomial of degree $4L+4$ in the variable $x$
which has to be zero.
This polynomial will be the subject of the next section.

To obtain eigenvectors corresponding to 
$\lambda=0$ one has to solve the boundary eqs.(\ref{leftb1}-\ref{rightb2})
using the bulk solution given by eq.(\ref{bulks}) with $x=\pm\i$.
 The calculation is given in the appendix.
It turns out that besides the  eigenvectors
 $(0,1,0,\ldots)$ and $(\ldots,0,1,0)$ of 
$M$, which are always present, one may have
additional eigenvectors for $\lambda=0$.
Running through the calculation it turns out that this happens if $\alpha_-\beta_+ + \alpha_+\beta_-=0$.
Under this condition  
 two further linearly independent solutions always 
exist.  
If all relevant boundary parameters vanish, i.\ e.\  
$\alpha_-=\alpha_+=\beta_-=\beta_+=0$ and if at the same time
$\alpha_z=-\beta_z$ for $L$ odd or $\alpha_z=\frac{1}{2\beta_z}$ for $L$ even
we have four additional solutions.
Two of them are just $(1,0,0,\ldots)$ and $(\ldots ,0,0,1)$.
Note that the degeneracy of the eigenvalue $\lambda=0$ might be higher than the
number of linearly independent eigenvectors since $M$ might be 
non--diagonalizable. This will be discussed in the appendix by considering
the polynomial equation which is given in the next section.

The calculations we have done so far enable us to give a complete set of
conditions under which  $M$ is non--diagonalizable. This is always the case if 
the degeneracy of an eigenvalue is higher than the number of linearly 
independent 
eigenvectors. The conditions for the eigenvalue
 $\lambda=0$ are derived in the appendix, whereas
the conditions for the eigenvalues $\lambda \neq 0$ are obtained from eq.(\ref{m}). 
$M$ is non--diagonalizable, if one of the following conditions is satisfied :
\begin{enumerate}
\item  $M$ has an eigenvalue $\lambda\neq 0$ which is more than twofold degenerate.
\item  $M$ has an eigenvalue $\lambda\neq 0$ which is twofold degenerate, but at least one 
       of the $\Omega_{ij}$ is different from zero.
\item  $\lambda=0$ is an eigenvalue of $M$, but it is more than sixfold degenerate.
\item  $\lambda=0$ is a sixfold degenerate eigenvalue of $M$, but one of the parameters 
       $\alpha_+,\alpha_-,\beta_+,\beta_-$ is different from zero.        
\item  $\lambda=0$ is a sixfold degenerate eigenvalue of $M$, but $\alpha_z\neq -\beta_z$ 
       for $L$ odd or $\alpha_z\neq\frac{1}{2\beta_z}$ for $L$ even respectively.
\end{enumerate}
If none of these conditions is satisfied, $M$ is diagonalizable.

\section{The polynomial equation}
Now we turn to the polynomial equation which determines the eigenvalues of $M$.
For later convenience, we define a new variable $z=x^2$.
As can be seen directly from $M$ (eq.\ (\ref{matrix_M}), 
one of the eigenvalues is
always zero, 
the others are obtained from eq.\ (\ref{x_lambda}) as
\be
\label{Lambda}
\Lambda_n=\frac{1}{4} (\sqrt{z_n}+\sqrt{1/z_n})
\ee
Since we have defined the $\Lambda_n$ to have a positive real part,
we will ---here and in the following--- always by definition take 
the square root which has positive real part.  The $z_n$ are the solutions of
the following polynomial which has been obtained
from eq.\ (\ref{detomega}) 
\ba
\label{pol}
p(z) & = & \frac{1}{(z-1)^2}\Bigl[z^{2L+4} - A \;(z^{2L+3}+z)
+(B+E^2)\;(z^{2L+2} +z^2) \nn\\ & &
+ (D+2E^2) \;(z^{2L+1}+z^3) +E^2 \;(z^{2L} +z^4)
-2E \; (z^{L+4}+z^L) \nn\\ & &+
\Bigl(\half\,(-1+A-B-D)- (-1)^{L} C - 2 E^2\Bigr) \;(z^{L+3}+z^{L+1})
\\ & & + \Bigl(-1+A-B-D+ 2 (-1)^L C+ 4 E -4 E^2\Bigr)\; z^{L+2} +1  \Bigr]\nn\\
& = & \frac{1}{(z-1)^2} q(z) = 0\nn
\ea
The coefficients are given by
\ba
\label{rates}
A & = & 2 (-1 + \a\g + \b\d + \k^2 + \rh^2)\nn\\
B & = & (-1 + 2 \a\g)(-1 + 2 \b\d) + 4 (-1 + \a\g) \rh^2+4 (-1 + \b\d) \k^2 
\nn\\
C & = &  (\a^2 \b^2+ \g^2 \d^2) \\
D & = & 2 (-1 + 2 \a\g) \rh^2 + 2 (-1 + 2 \b\d) \k^2 \nn\\
E & = & 2 \k \rh \nn
\ea
 
Note that the polynomial $p(z)$
is already completely determined by five complex
parameters although we started with six parameters in the original Hamiltonian.
This can be explained by the existence of a similarity transformation 
of the form
\be
\label{trafo}
H^{\prime } = U H U^{-1} \;\;\mbox{with}\;\;
U = \prod_{j=1}^{L} I_1\otimes\cdots I_{j-1}\otimes
\left(
\begin{array}{ll}
        1 & 0 \\
        0 &\epsilon
\end{array}
 \right)\otimes I_{j+1}\otimes\cdots\otimes I_L
\ee
containing one free parameter $\epsilon$. Here $I_j$ stands for the 
identity matrix at the site $j$. By using this similarity 
transformation, 
the four boundary parameters $\a, \g, \d$ and $\b$
are transformed as follows
\be
\a \rightarrow \epsilon \: \a; \;\;\; \g \rightarrow 1/\epsilon \: \g; \;\;\;
\d \rightarrow \epsilon \: \d; \;\;\; \b  \rightarrow 1/\epsilon \: \b 
\ee
and by choosing a particular value of $\epsilon$, one can always fix
one of the boundary parameters.

The polynomial $q(z)$
has a very special form, because in comparison with a general
polynomial of degree $2L+4$ many of the coefficients
are zero. This changes of course when it is divided 
by $(z-1)^2$.

Observe that the polynomial $p(z)$ has degree $2 L+2$ although  
the diagonal form of $H_{long}$ given by eq.\ (\ref{spectrum}) 
has only $L+2$ fermionic 
excitations $\Lambda_n$.
The reason therefore is the quadratic relation between
$z$ and $\Lambda$. Since with $z$ also $\frac{1}{z}$ is a solution of the
polynomial, one gets each value of $\Lambda$ twice. Taking half of them
and adding the additional
eigenvalue $0$ which was already mentioned in section 2
and explicitly excluded from the set of
solutions in section 3
gives exactly the $L+2$ fermionic excitations.

Special solutions of this polynomial will be studied in the next section.

\section{Factorization of the polynomial in cyclotomic polynomials}
The study of the factorization properties of the polynomial given by
eq.\ (\ref{pol}) 
represents  a very interesting
mathematical problem. Furthermore,
factorizations of the polynomial into cyclotomic polynomials 
which we are going to present below are very important because they allow
to calculate the whole 
spectrum and other properties of $H_{\mbox{\scriptsize{long}}}$ 
analytically. 

For some special choices of the parameters $A$, $B$, $C$, $D$ and $E$
the polynomial factorizes exactly into cyclotomic polynomials. 
These factorizations were found  using the following algorithm:
A factorized polynomial of degree $2L+4$ of the form
\be
\label{factored_polynomial}
f(z) = \left[ \prod_{i=1}^{k} (1-p_i \, z^{n_i}) \right] (1-p_{k+1} z^{2L+4-
\sum n_i})
\ee
is expanded for a fixed value of $k$ (corresponding to a fixed 
number of factors), a fixed value of $L$ and for all possible 
combinations of the $n_i$. The coefficients of
the expanded polynomial $f(z)$ which are functions of the $p_i, i=1,\ldots,
k+1$ are compared to the coefficients of the original polynomial $q(z)$. 
In this way, one obtains a set of $2L+4$ coupled equations for the $p_i$ and
the coefficients $A,B,C,D$ and $E$ which has been solved using 
{\it Maple}. 
Typically values of $L=5,6,7$ were used; for smaller values of $L$ 
the equations do not reflect the general situation because some
exponents coincide. For larger $L$ however, the number of partitions
of $2L+4$ into the $n_i, i=1,\dots,k$
gets too large. Among the solutions only those were kept which are valid
for arbitrary $L$ and not only for the special $L$ used in the calculation.
$k$ has been varied between 1 and 4 (i.\ e.\ factorizations in $2,3,\ldots,5$
factors were studied). For $k=5$ (six factors) the program did not run properly
-- it needed too much memory. However, 
$q(z)$ cannot factorize into a a larger number of factors of the above
form (\ref{factored_polynomial}) than six
with the condition that the corresponding $n_i$
appear explicitly as exponents in $q(z)$. 
In this case the only possible combination for the $n_i$ would be 
\be 
\label{choice}
n_1 = n_2 = L\,;\;\; n_3 = n_4 =
n_5 =n_6 = 1
\ee

In table I  all factorizations which were calculated as described above
are listed. The factorizations in six factors (entries 14 -- 16)
were found by solving the system of coupled equations for the choice
(\ref{choice}) of the $n_i$ and various
choices of the $p_i$. 
Therefore the list might not be 
exhaustive for the factorizations into six factors.

\begin{table}
\caption{Cases where the polynomial factorizes into cyclotomic polynomials}
\begin{center}
\lineup
\footnotesize
\begin{tabular}{@{}lrrrrr|ll}
\br
case &   $A$ & $B$ & $(-1)^L C$ & $D$ & $E$ & $q(z)$ \\ \mr \\ 
1.) & 1 & 0 & 0 & 0 & 0 & $(1-z)   (1-z^{2L+3})$ \\
2.) & 0 & -1& 0 & 0 & 0 & $(1-z^2) (1-z^{2L+2})$ \\
3.) &-1 & 0 & 0 & 0 & 0 & $(1+z)   (1-z^{L+1})(1-z^{L+2})$\\
4.) &1 & -1 & 0 & 1 & 0 & $(1+z) (1-z)^2(1+z^{2L+1})$ \\
5.) &0 &  0 & $\frac{1}{2}$ & 0 & 0 & $(1-z^{L+1})(1-z^{L+3})$ \\
6.) &0 &  0 & $-\frac{1}{2}$ & 0 & 0 & $(1-z^{L+2})^2$ \\
7.) &1 &  0 & 1 & 0 & 0 & $(1-z) (1-z^{L+1}) (1+z^{L+2})$ \\
8.) &1 &  0 & -1 & 0 & 0 & $(1-z) (1+z^{L+1}) (1-z^{L+2})$ \\ \mr
9.)& 2 & 1 & $s+1/s$ & 0 & 0 & $(1-z)^2(1-sz^{L+1})(1-1/s\,
z^{L+1})$ \\ 
10.) & $s+1/s$ & 1& $\frac{1}{2}(2+s+1/s)$ & 0 & 0&
$(1-s z)(1-1/s \,z)(1-z^{L+1})^2$ \\
11.)& $1+s+1/s$ & $1+s+1/s$ & 0 & -1 & 0 & $(1-sz)
(1-1/s\, z)(1-z)(1-z^{2L+1})$ \\ 
12.)& 0& $-1-s-1/s$ & $\frac{1}{2}(-2+s+1/s)$ & -2 & 1 & 
$(1-sz^2) (1-1/s \,z^2)(1-z^{L})^2$ \\ 
13.)& $s+1/s$ & 1 & $-2-s-1/s$ & 
$-2-s-1/s$ & 1 & $(1-sz) (1-1/s \, z)(1+z^2)(1-z^{L})^2$ \\ 
14.)& $2+s+1/s$ & $1+2s+2/s$ & $4+2s+2/s$ & 
$-4-s-1/s$ & -1 & $(1-sz) (1-1/s \,z)(1-z)^2(1+z^{L})^2$ \\ 
15.)& $2+s+1/s$ & $1+2s+2/s$ & $-4-2s-2/s$ & 
$-4-s-1/s$ & 1 & $(1-sz) (1-1/s \,z)(1-z)^2(1-z^{L})^2$ \\ 
16.)& $-2+s+1/s$ & $1-2s-2/s$ & 0 & $-2-s-1/s$ & 
 1 & $(1-sz) (1-1/s \,z)(1+z)^2(1-z^{L})^2$ \\ 
\mr \br
\end{tabular}
\end{center}
\end{table}

The entries $10$ to $16$ 
each furnish a one-parameter family of solutions (where the parameter
is called $s$) for which 
a factorization in cyclotomic polynomials appears. 
Notice however that $E$ is the only parameter
whose value is always fixed, independently of $s$. 
Since $E= \alpha_z \beta_z$ this means that
in all  cases presented in table I the product of the coefficients in front
of the diagonal boundary terms is always fixed. 
Moreover, the entries $10$ to $16$  provide
examples where some of the zeros of the polynomial $q(z)$
are always independent of $L$ (e.\ g.\ $z = s$ or $z = 1/s$). 
We will come back to this point in section 7 and in the discussion
(section 14).

Looking at the entries $10$ to $16$ the remark we made
at the end section 3 that in special cases
the Hamiltonian might not be diagonalizable becomes clear:
It is  possible to choose the parameter $s$ as a function of $L$ 
in such a way that the polynomial has
zeros which are more than twofold degenerate. 
Take for example the case $13$ and choose $s$ equal to one
solution of $1-z^L = 0$. Then the corresponding zero of $q(z)$ is
threefold degenerate for the value of $L$ chosen above.
In this case, one can not
find more than two independent eigenvectors for $M$ (cf. eq.\ (\ref{m}))
for the corresponding degenerate eigenvalue. 
Thus in these special cases (which can be constructed analogously for
the other entries $10$ to $16$) 
the matrix $M$ is not diagonalizable.

For all examples in table I it is possible to calculate 
the spectrum and the ground state energy exactly.
In the following sections we will give the explicit expressions
for the ground state energies and for some excitations.
With the insight gained from these
exactly solvable cases we will later also treat the general
case in the limit of large $L$ \cite{paper2}.

\section{Examples of exact calculations of spectra for the finite lattice}
Let us now present two examples how to calculate the spectrum of
$H_{long}$ from the factorized
form of the polynomial. We will first take case 4 from table I. 
The factorized form of the polynomial $q(z)$ as given in table I suggests
the ansatz $z=\exp( \i \pi + \frac{2 \i \phi}{2L+1})$.
This leads to  the solutions
\be
\phi = n \pi, \;\;\;\;\; n = 1,\ldots L
\ee 
The factor $(1+z)$ leads to the additional solution $z=-1$  which means
$\Lambda=0$. The factor $(1-z)^2$ has to be divided out because the 
fermionic eigenvalues are given by the zeros of $p(z)$ and not of
the ones of $q(z)$ which is given in table I (also cf.\ eq.\ (\ref{pol})).

Then the energies of the
fermionic excitations are given by $\Lambda = 0$ (twice) and, using
eq.\ (\ref{Lambda}), by
\be
\label{energy1.}
\Lambda_n = \frac{1}{2} \sin \left( \frac{n \pi}{2L+1} \right)
\;\;\;\;\; n = 1,\ldots L 
\ee

The energies of the fermionic excitations in the other cases from table I
have a similar form.
However, in the cases 10, 11 and 13 to 16 there is always one solution
with roots $z=s^{\pm 1}$ leading to a fermionic energy 
$\Lambda = \frac{1}{4} (\sqrt{s}+\sqrt{1/s})$ which is ---
in contrast to the fermionic energies 
obtained in eq.\ (\ref{energy1.}) --- independent 
of the lattice length $L$.
We will see later \cite{paper2} that this energy can be connected
to a boundary bound state.
The nature of this state will be elucidated by studying
the corresponding spatial profiles and by comparing some spatial
profiles  for  special choices 
of the parameters to a Bethe-Ansatz solution of the $XX$--chain with
only diagonal boundary terms \cite{Skorik_Saleur}. 
This will be described in detail
in \cite{paper2}.

We now consider the case 9 of table I which
is special because the roots of the polynomial all depend
on a parameter $s$ and in general do not lie on the unit circle. Therefore
we will briefly present the corresponding results here.
For $L$ even we obtain the solutions $\Lambda=0$ and 
\ba
\label{excitations_10}
\Lambda_n &=& \frac{1}{2} \sin \left( \frac{(2n+1)\pi}{2(L+1)} 
+ \frac{\i \ln{s}}
{2(L+1)} \right) \;\;\;\;\; n = 0,1, \ldots \frac{L}{2}\nn\\
\Lambda_n &=& \frac{1}{2} \sin \left( \frac{(2n+1)\pi}{2(L+1)} 
- \frac{\i \ln{s}}
{2(L+1)} \right) \;\;\;\;\; n = 0,1, \ldots \frac{L-2}{2}
\ea
for $|Im(\ln{s})| \leq \pi$. 

For $L$ odd we have accordingly $\Lambda=0$, and
\ba
\label{excitations_10_odd}
\Lambda_n &=& \frac{1}{2} \sin \left( \frac{n\pi}{L+1} + \frac{\i \ln{s}}
{2(L+1)} \right) \;\;\;\;\; n = 1, \ldots \frac{L+1}{2}\nn\\
\Lambda_n &=& \frac{1}{2} \sin \left( \frac{n\pi}{L+1} - \frac{\i \ln{s}}
{2(L+1)} \right) \;\;\;\;\; n = 0,1, \ldots \frac{L-1}{2}
\ea
for $0 \leq Im(\ln{s}) \leq \pi$. For $-\pi \leq Im(\ln{s}) \leq 0$
one has to interchange the limits of $n$ in the two sets of eigenvalues
given by eq.\ (\ref{excitations_10_odd}).

In this example,  we can see explicitly how the parameter $s$ appears in
the spectrum. The argument of the sine is shifted
by the $s$--dependent term $\frac{\i \ln{s}}{2L+2}$.

For the examples given in table I, it is also possible to calculate 
the ground state energies exactly.
The corresponding expressions will be given
in the next section.

\section{Exact expressions for the ground state energies of $H_{long}$}
Let us first consider the case 4 of table I again. 
The ground state energy is given by summing up all negative
eigenvalues of $M$ (cf.\ eq.\ (\ref{spectrum})), this leads to
\be
E_0 = -\frac{1}{2} \sum_{n=1}^L
\sin \left( \frac{n \pi}{2L+1} \right)
= - \frac{1}{4} \cot {\frac{\pi}{4L+2}}
\ee

In the case 9, also discussed in section 6,
the ground state energy for $L$ even is given by
\ba
\label{gs1}
\fl
E_0 &=& -\frac{1}{2} \sum_{n=0}^{(L-2)/2} \left[\sin \left( \frac{(2n+1)\pi}
{2(L+1)} + \frac{\i \ln{s}}{2(L+1)} \right) +\sin \left(\
\frac{(2n+1)\pi}{2(L+1)}
 - \frac{\i \ln{s}}{2(L+1)} \right) \right]\nn\\
\fl
& &  - \frac{1}{2} \cosh \left(
\frac{ \ln{s}}{2L+2} \right)
 =  -\half \frac{\cosh{\frac{\ln(s)}{2L+2}}}{\sin{\frac{\pi}{2L+2}}}
\ea
For $L$ odd we obtain
\ba
\label{gs2}
E_0 &=& - \half \left[\cot{\frac{\pi}{2L+2}} \cosh{\frac{\ln(s)}{2L+2}}
+ \i \; \sinh{\frac{\ln(s)}{2L+2}}\right]\nn\\
&=& -\half \frac{\cosh{\frac{\ln(s)+ \i \pi}{2L+2}}}{\sin{\frac{\pi}{2L+2}}}
\ea

This expression is indeed real if the original Hamiltonian is chosen to be
hermitian.
This can be seen by imposing hermitian boundary terms in the Hamiltonian
$\a = \g^*, \b = \d^*$ and solving the system of equations
\be
\begin{array}{ccccc}
A &=& 2   &=& 2 (-1+ |\a|^2+ |\b|^2),\;\;\;\;\;\nn\\
B &=& 1   &=&  (1-2 |\a|^2)( 1-2 |\b|^2),\;\;\;\\
(-1)^L C &=& s + \frac{1}{s} &=& 2 |\a|^2 |\b|^2 \cos{2(\xi_{\alpha} +
\xi_{\beta})}\nn
\end{array}
\ee
where $\xi_{\alpha},\xi_{\beta}$ are the phases of $\a$ and $\b$, respectively.
These equations have only the solution $|\a|=|\b|=1,
s= \exp\left(\pm 2 \i (\xi_{\alpha} + \xi_{\beta})\right)$.
For this
choice of $s$
the ground state energies given by eqs.\ (\ref{gs1}) and (\ref{gs2})
are always real.

In table II, the expressions for the ground state energies for all 
cases from table I are listed. Again, they correspond to $H_{long}$.
In section 12, we will describe how the corresponding ground state energies 
for the original Hamiltonian $H$ are obtained from the ones for
$H_{long}$.
It is remarkable that despite the non--diagonal boundary terms
all expressions for the ground state energies of $H_{long}$ 
are given in terms of trigonometric functions. The only exception is case 
10 where hyperbolic functions appear in the expression for the ground state
energy (see eqs.\ (\ref{gs1}) and (\ref{gs2})). However, even in this
case the model is integrable.

The virtue of table II is that one can explicitly see
how the ground state energy is changing with different boundary parameters
(we refer to table III of section 12 for some examples in
which boundary parameters
of the Hamiltonian correspond to a given choice of the $A,B,C,D$ and $E$).
This is especially interesting when studying the thermodynamic limit
as we will do in \cite{paper3}. There we will show that the Hamiltonian
with arbitrary boundary terms corresponds to a  conformal invariant
theory.  In particular, if one expands the expressions of table II
in powers of $\frac{1}{L}$, one can already see
that they have the typical form 
of the ground state expansion corresponding to
 a conformally invariant theory and
one can directly read off the conformal charge and the surface free
energy for the different boundary parameters.
 
Notice that the table includes  a well--known spin chain: 
the XX--chain with open boundaries is
case 10 with $s=-1$.

The $s$--dependent cases 9 -- 16 are
of special interest because the $s$-dependence is still manifest in the
expressions of the ground state energy and one can see
how this family of ground states  varies with $s$. 
Note that in all cases 10 -- 16
the $s$-dependent terms appear additively to the $L$-dependent part
of the ground state energy, (and therewith contributes additively to the 
surface free energy in the expansion for $L \rightarrow \infty$).
The physical consequences of the $L$--independent solutions
will be discussed in the next paper in connection 
with  boundary states \cite{paper2}.

In case 9 however, there is no such simple structure
and the parameter $s$ is coupled with $L$
as argument of the $\cosh$-term. Indeed the case 9 is special
as can be seen already from table I (the $s$-dependence appears
in the factors of the polynomial which have degree $L+1$, and not
in the factors of degree $1$ or $2$ as in all other $s$-dependent cases).
This case will be discussed in detail in the following two articles
\cite{paper2,paper3}.

\begin{table}
\caption{Ground state energy of $H_{long}$ for the cases from table I}
\lineup
\footnotesize
\begin{tabular}{@{}l|ll}
\br
example &   $L$ even and $L$ odd\\ \mr
1.) & $\frac{1}{4}-\frac{1}{4}\frac{1}{\sin{\frac{\pi}{4L+6}}}$ &  \\
2.) & $\frac{1}{4}-\frac{1}{4}(\frac{1}{\sin{\frac{\pi}{2L+2}}}+
\cot{\frac{\pi}{2L+2}})$ & \\
4.) & $-\frac{1}{4} \cot{\frac{\pi}{4L+2}}$ &\\ 
11.) & $\frac{1}{4} -\frac{1}{4} \frac{1}{\sin{\frac{\pi}{4L+2}}} -
\frac{1}{4} (s^{1/2}+s^{-1/2})$ & \\\mr
 & $L$ even &  $L$ odd\\ \mr
3.) & $\half - \frac{1}{4} (\frac{1}{\sin{\frac{\pi}{2L+2}}}+\cot
{\frac{\pi}{2L+4}})$ & 
$\half - \frac{1}{4} (\frac{1}{\sin{\frac{\pi}{2L+4}}}+\cot{\frac
{\pi}{2L+2}})$ \\
5.)& $\half - \frac{1}{4} (\frac{1}{\sin{\frac{\pi}{2L+6}}}+\frac{1}{\sin
{\frac{\pi}{2L+2}}})$ &  
$\half - \frac{1}{4}(\cot{\frac{\pi}{2L+2}}+\cot{\frac{\pi}{2L+6}})$\\
6.)& $\half-\half\cot{\frac{\pi}{2L+4}}$ & $\half -\half\frac{1}{\sin{
\frac{\pi}{2L+4}}} $  \\
7.)& $\frac{1}{4} -\frac{1}{4}(\frac{1}{\sin{\frac{\pi}{2L+2}}}+\frac{1}
{\sin{\frac{\pi}{2L+4}}})$ & 
$\frac{1}{4} -\frac{1}{4}(\cot{\frac{\pi}{2L+2}}+\cot{\frac{\pi}{2L+4}})$\\
8.)& $\frac{1}{4}-\frac{1}{4}(\cot{\frac{\pi}{2L+2}}+\cot{\frac{\pi}{2L+4}})$
& $\frac{1}{4} -\frac{1}{4}(\frac{1}{\sin{\frac{\pi}{2L+2}}}+
\frac{1}{\sin{\frac{\pi}{2L+4}}})$ \\ \mr
9.) & $- \half \frac{\cosh{\frac{\ln(s)}{2L+2}}}{\sin{\frac{\pi}{2L+2}}}$&
$-\half \frac{\cosh{\frac{\ln(s)+ \i \pi}{2L+2}}}{\sin{\frac{\pi}{2L+2}}} $ \\
10.)& $\half -\half \frac{1}{\sin{\frac{\pi}{2L+2}}}-\frac{1}{4} 
(s^{1/2}+s^{-1/2})$ &
$\half -\half \cot{\frac{\pi}{2L+2}}-\frac{1}{4} (s^{1/2}+s^{-1/2})$ \\
12.) & $\half -\half \cot{\frac{\pi}{2L}}-\frac{1}{4} (s^{1/4}+s^{-1/4})
-\frac{\i}{4} (s^{1/4}-s^{-1/4})$ & 
$\half -\half \frac{1}{\sin{\frac{\pi}{2L}}}-\frac{1}{4} (s^{1/4}+s^{-1/4})
-\frac{\i}{4} (s^{1/4}-s^{-1/4})$\\
13.) & $\half-\frac{1}{2 \sqrt{2}} -\half \cot{\frac{\pi}{2L}}-\frac{1}{4} 
(s^{1/2}+s^{-1/2})
$ &
$\half-\frac{1}{2 \sqrt{2}} -\half \frac{1}{\sin{\frac{\pi}{2L}}}-\frac{1}{4} 
(s^{1/2}+s^{-1/2}) $\\
14.) & $-\half \frac{1}{\sin{\frac{\pi}{2L}}}-\frac{1}{4} (s^{1/2}+s^{-1/2})$ &
$-\half \cot{\frac{\pi}{2L}}-\frac{1}{4} (s^{1/2}+s^{-1/2})$ \\
15.) & $ -\half \cot{\frac{\pi}{2L}}-\frac{1}{4} (s^{1/2}+s^{-1/2})$ &
$ -\half \frac{1}{\sin{\frac{\pi}{2L}}}-\frac{1}{4} (s^{1/2}+s^{-1/2})$\\
16.) & $\half -\half \cot{\frac{\pi}{2L}}-\frac{1}{4} (s^{1/2}+s^{-1/2})$ &
$\half -\half \frac{1}{\sin{\frac{\pi}{2L}}}-\frac{1}{4} (s^{1/2}+s^{-1/2})
$\\\br\br

\end{tabular}

\end{table}


\section{Example of a spectrum of $H_{long}$ with asymmetric
bulk terms}

Up to now, we did not look explicitly at Hamiltonians with
asymmetric bulk terms which we already mentioned in the introduction.  
Recall that we can map such a Hamiltonian $\tilde{H}$
given by eqs.\ (\ref{Ham_nh})
to a Hamiltonian $H$ of the form given by eq.\ (\ref{Ham})
which has symmetric bulk terms and $L$--dependent boundary terms
of the special form given by eq.\ (\ref{trans}). Using the 
methods previously described we can solve the eigenvalue problem
for the  Hamiltonian $H_{long}$ and in this way 
obtain the spectrum of $\tilde{H}_{long}$ where $\tilde{H}_{long}$
is obtained from $\tilde{H}$ in the same way as $H_{long}$ from $H$
by adding one site at each end of the chain. 
We will carry this out for
one example. We choose a Hamiltonian $\tilde{H}_{long}$ whose transformed
$H_{long}$ has boundary terms corresponding to the case 9.
Notice that this is the only factorizable case where this can be done
independently of $L$ if $p \neq q$, 
i.\ e.\ the analysis can simultanously be carried
out for all Hamiltonians $\tilde{H}$ of the chosen type with arbitrary
length. In all other cases starting from one factorizable case and
changing the length $L$ would result in boundary parameters belonging
to another (perhaps not even factorizable) case.

One possible choice for the boundary parameters in the case 9 is 
given by  (as can be directly seen from eq.\ (\ref{rates})):
$$ \a \g =1 ;\;\;\b \d =1;\;\; 
(-1)^L (\a^2 \b^2+ \g^2 \d^2) = 
s + \frac{1}{s}; \;\;\k=0=\rh
$$

Expressing $\a$ in terms of $\g$ and $\d$ in terms 
of $\b$ and using eq.\ (\ref{trans}), we obtain
$s=(-1)^L \left(\frac{\b^{\prime}}{\g^{\prime}}\right)^2 Q^{2-2L}$
(or $1/s = (-1)^L \left(\frac{\b^{\prime}}{\g^{\prime}}\right)^2 Q^{2-2L}$). 
Using the results obtained in eq.\ (\ref{excitations_10}) 
we get for the fermionic 
excitations for even $L$
\ba
\label{energy}
\Lambda &=&  \half \sin \left( \frac{(2n+1) \pi}{2L+2} \pm
\i \left[\frac{2 \ln{Q} +\ln{\frac{\b^{\prime}}{\g^{\prime}}}}{L+1} - 
\ln{Q}\right] \right)
\;\;\; n=0,\dots,\frac{L-2}{2}\nn\\
\Lambda &=&  \half \cosh \left[\frac{2 \ln{Q} +
\ln{\frac{\b^{\prime}}{\g^{\prime}}}}{L+1} - \ln{Q}\right]
\ea
The quasi-momenta $\frac{(2n+1) \pi}{2L+2}$ are shifted by 
the constant $\i \left[\frac{2 \ln{Q} +
\ln{\frac{\b^{\prime}}{\g^{\prime}}}}{L+1} -
\ln{Q}\right]$. The second $L$--independent term of this constant
is typical for
asymmetric bulk terms as will be seen in \cite{paper3}.
A similar expression has been obtained in \cite{toy_paper}
for the fermionic excitations of $\tilde{H}$ with totally asymmetric
bulk terms ($p=1, q=0$), $\a \neq 0$, $\b \neq 0$ and
  $\g = \d = \az = \bz = 0$.

The ground state energy is given by summing up all negative
eigenvalues of $M$, leading to
\be
E_0 = - \frac{\cosh(\frac{\ln\left(\frac{q \b^{\prime}}{p \g^{\prime}}\right)}
{L+1}-
\ln{Q})}{2 \sin(\frac{\pi}{2L+2})}
\ee
Observe that for $\frac{p}{q}=1$ we obtain the previous expression 
calculated for the case 9.

\section{Projection method and the $\sigma^x$ 1-point functions}

Up to now we have dealt only with the Hamiltonian $H_{long}$ 
given by eq. (2.1) which was
obtained from $H$ (see eq. (1.2)) by adding one lattice site at
 each end of the chain.
In this section we will explain how the spectrum of $H$ is
related to the spectrum of $H_{long}$.

Since, as mentioned in section 2, $\sigma^x_0$ and $\sigma^x_{L+1}$
 commute with $H_{long}$ the
spectrum decomposes into four sectors $(++,+-,-+,--)$ corresponding
to the eigenvalues $\pm1$ of $\sigma^x_0$ and $\sigma^x_{L+1}$.
The eigenvalues and eigenvectors of $H$ are related to the $(++)$-sector
in the following way:

If $|E\ket$ is an eigenvector of $H$ corresponding to an eigenvalue $E$
then $|E_{long}\ket=|+\ket\otimes|E\ket\otimes|+\ket$ is an eigenvector of $H_{long}$
corresponding to the same eigenvalue $E$, where $\sigma_0^x|+\ket=|+\ket$
and $|E_{long}\ket$ is element of the space $\BCT^2 \otimes \BCT^L \otimes \BCT^2$.
Thus the whole spectrum
of $H$ is contained in the spectrum of $H_{long}$ projected onto 
the $(++)$-sector.
Since the dimension of the of $(++)$-sector is $2^L$
 we conclude that the spectrum of $H$ is identical to the spectrum
of $H_{long}$ projected onto the $(++)$-sector. 

Before describing how we will proceed to project to the $(++)$-sector,
we make some definitions which are
needed later.
First, we want to remind the reader that $M$ defined by eq.(\ref{matrix_M}) 
always has a twofold
degenerate eigenvalue $0$ corresponding to eigenvectors $(0,1,0,0,\cdots)$
and $(0,0,\cdots,1,0)$ of $M$. Using the Clifford operators $\tau^{\pm}_j$ given by
eq.(\ref{tau}), we now define the corresponding ladder operators  
\be
b_0=(\tau_0^+-\i\tau_{L+1}^-)/2 \quad ,\quad
a_0=(\tau_0^+ +\i\tau_{L+1}^-)/2 \quad .
\label{zeroladderop}
\ee
We also define the vacuum representation with the
lowest weight vector $|vac\ket$ by
\be
a_k |vac\ket=0 \quad \forall k \quad .
\ee
Because we are interested in eigenstates of $\sigma_0^x$ and $\sigma_{L+1}^x$
we define the vectors 
\be
|v^{\pm}\ket=\frac{1}{\sqrt{2}}(|vac\ket\pm|0\ket) \quad ,
\label{vplusminus}
\ee
where $|0\ket=b_0|vac\ket$. Observe that 
\be
\sigma_0^x |v^{\pm}\ket = \pm |v^{\pm}\ket \quad .
\label{vpm}
\ee

Now we will proceed in three steps.
In the first step (subsection 9.1)
we will show using some algebraic considerations that
the vectors $|v^{\pm}\ket$ are also eigenvectors of $\sigma_{L+1}^x$. 
It will turn out that the eigenvalues of $\sigma_{L+1}^x$ corresponding to
the eigenvectors $|v^{+}\ket$ and $|v^{-}\ket$ always have opposite signs, i.e. 
\be
\sigma_{L+1}^x|v^{\pm}\ket=\pm\eta |v^{\pm}\ket  \label{eta}
\ee
with $\eta^2=1$. The value of $\eta$ plays a crucial role in the following.
We will also show that
 the $(++)$-sector consists either of the states 
\be
\prod_{j=1}^r b_{k_j}|v^+\ket \quad \mbox{with $r$ even}
\label{even}
\ee
or
\be
\prod_{j=1}^r b_{k_j}|v^-\ket \quad \mbox{with $r$ odd}
\label{odd}
\quad ,
\ee
where $0 < k_j \leq k_{j+1}$. Since 
$k_j\neq 0$, the creation operator
of the spurious zero mode defined by eq.(\ref{zeroladderop}) does not appear 
in (\ref{even}) and (\ref{odd}).
The groundstate of $H$ corresponds to $|v^+\ket$ or to 
$b_k |v^-\ket$, where $b_k$ denotes the creation
operator corresponding to the fermionic energy with smallest real part 
and $k\neq 0$. 

Whether the $(++)$-sector consists of the states (\ref{even}) or (\ref{odd}) 
will be shown to depend 
on the value of $\eta$ (see eq.(\ref{eta})) which will
be calculated by
computing the expectation
value $\bra v^+|\sigma^x_{L+1}|v^+\ket$.
Note that we define 
$\bra v^{\pm} |$ via 
\be
\bra v^{\pm} | = \frac{1}{\sqrt{2}}(\bra vac | \pm \bra vac | a_0)  \quad ,
\label{bra_vpm}
\ee 
where $\bra vac|$ denotes the
left vacuum of $H_{long}$, i.e.
\be
\bra vac|b_k=0 \quad .
\label{leftvac}
\ee
Note that if $H_{long}$ is not hermitian,  $\bra vac|$ is not equal to
$| vac \ket^{\dagger}$
in general.

In the second step (section 10), we will show 
how to calculate the one--point function
\be
f(j)=\bra v^+|\sigma^x_j|v^+\ket   
\ee
In the present context, this is done merely for technical reasons
in order to calculate $f(L+1)$, however it will be essential
in another context. Namely, 
the explicit
calculation of $f(j)$ will be presented in the following paper \cite{paper2}.
We would already like to remark that the calculation of $f(j)$ is similar to
 the calculation of the two-point function
\be
g(i,j)=\bra v^+|\sigma^x_i\sigma^x_j|v^+\ket \quad .
\ee
Note that due to eq.(\ref{vpm}) one already sees that
$f(j)=g(0,j)$. We show in section 10
 that the functions $f(j)$ and $g(i,j)$ are both given
by Pfaffians of submatrices of the same matrix. We will furthermore
generalize these considerations
to states of the form (\ref{even}) and (\ref{odd}). 
These results also apply to $H$ although we started with the
larger space of states of $H_{long}$.

Using eq.(\ref{crucial}) and eq.(\ref{crucial2}), 
we will obtain determinant representations
for $f(j)$ and $g(i,j)$ which can be treated analytically in the calculation of $f(L+1)$.
This calculation can be found in section 11 and will be the third step. 

\subsection{Algebraic considerations}
In this subsection, we will show that the $(++)$-sector consists either 
of the states given by (\ref{even}) or (\ref{odd}) and clarify the role 
of $\eta$ given in eq.(\ref{eta}). Because 
\be
(\phi^{\pm}_k)^+_0=(\phi^{\pm}_k)^-_{L+1}=0, \forall k\neq 0 
\label{0komponenten}
\ee
which can
be seen directly from the matrix $M$, we obtain 
the commutation relations      
\be
[\sigma^x_{L+1},b_k]=[\sigma^x_{L+1},a_k]=
\{ \sigma^x_0,b_k \}=\{ \sigma^x_0,a_k\} =0
\quad , \forall
k\neq 0 
\label{komsiga}
\ee
from eqs.(\ref{a_dagger}) and (\ref{a})
and therefore 
\be
[\sigma^x_0,N_k]=[\sigma^x_{L+1},N_k]=0 \quad , \forall k\neq 0 \quad .
\label{komsign}
\ee
Due to eq.(\ref{komsign}) the vectors $|v^{\pm}\ket$ have to
be eigenvectors of $\sigma_{L+1}^x$, i.e.
\be
\sigma_{L+1}^x|v^{\pm}\ket=\eta^{\pm} |v^{\pm}\ket \quad. 
\label{etapm}
\ee
Thus the sector containing $|v^+\ket$ respectively $|v^-\ket$ 
 is well defined and given by the value of $\eta^+$ and $\eta^-$
respectively. Note that eq.(\ref{etapm}) is not as precise as
eq.(\ref{eta}) because eq.(\ref{eta}) implies $\eta^+=-\eta^-$.

Due to eq.(\ref{komsiga}) we can make the following statement concerning
the vectors of a given sector: If an arbitrary vector $|v\ket$ is element
 of the $(\pm \; \epsilon)$-sector where $\epsilon \in \{+,-\}$ 
then $b_k|v\ket$ with $k\neq 0$ is element of
the $(\mp \; \epsilon)$-sector.
Now one has to distinguish two cases: 

First, if $|v^+\ket$ is an element of the $(++)$-sector, i.e. $\eta^+=+1$, 
 then all the states
given by (\ref{even}) are also elements of the $(++)$-sector. 
The vector $|v^-\ket$  then has to be an element of the $(--)$-sector
because otherwise the $(--)$-sector would be missing in the space of
states which is not the case. Thus we have $\eta^-=-1$.

Second, if $|v^+\ket$ is an element of the $(+-)$-sector, i.e. $\eta^+=-1$,
 then $|v^-\ket$ 
has to be an element of the $(-+)$-sector.
Otherwise $|v^-\ket$ would be an element of the $(--)$-sector  
and there would be no $(++)$-sector. As a consequence we have that
$\eta^-=+1$
and all the states given by (\ref{odd}) are elements of the $(++)$-sector.

In both cases, the subspace spanned by the vectors (\ref{even}) respectively
(\ref{odd})
has dimension $2^L$  and thus they form a basis of the $(++)$-sector.
Because the values of $\eta^+$ and $\eta^-$ always have  
opposite signs, we will use the variable $\eta$ defined
by eq.(\ref{eta}) in the following.

\begin{section}{1- and 2-point functions of $\sigma^x$}
In this section show how to compute the
1- and 2-point functions for the $\sigma^x_j$ operator
with respect to 
the states $|v^{\pm}\ket$ defined by eq.(\ref{vplusminus}) 
and $\bra v^{\pm}|$ defined by eq.(\ref{bra_vpm})
following the way of Lieb, Schultz and Mattis (LSM) \cite{LSM}.
The computation of the correlators of the operators $\sigma^y_j$ 
can be done similarly.
In this paper, however, 
 we are only interested in the value of $\bra v^+|\sigma_j^x|v^+\ket$ 
at the particular point $j=L+1$
which is calculated in section 11 using the results of this section. Due to
eq.(\ref{eta}) we thereby obtain the value of $\eta$ which is needed for the
projection from $H_{long}$ to $H$ as described in the previous section. 
The calculation for general values of $j$ will be part of the second paper 
of this series \cite{paper2}.

The main difference in comparison to the problem of LSM \cite{LSM}
is that in our case the 1-point functions
do not vanish because of the non-diagonal boundary terms we are considering.
They can be calculated in the same way as the 2-point correlators.
At the end of this section, we will briefly remark  how to compute 
the 1- and 2-point functions  for excited states of $H_{long}$ of the form (\ref{even}) and
(\ref{odd}). This is not necessary for the calculation of $\eta$ defined
in eq.(\ref{eta}), but it is needed
for the calculation of the 1- and 2-point functions 
for the eigenstates of $H$. Note that if $\eta=-1$ the ground state of $H$
corresponds to an excited state of $H_{long}$. 

We now proceed to the calculation of the 1-- and 2--point functions.
Writing $\sigma_j^x$ in terms of the $\tau^{\pm}_i$ defined in eq.(\ref{tau}) we obtain
\be
\bra v^{\pm} \mid \sigma^x_j \mid v^{\pm} \ket = \pm
 (-\i)^j \bra v^{\pm} \mid \tau^-_0 \prod_{k<j} \tau_k^+\tau_k^-  \tau_j^+ \mid v^{\pm} \ket 
\quad ,
\label{fjintau}
\ee
which is up to the sign exactly the 2-point function 
$\bra v^{\pm} \mid \sigma_0^x \sigma_j^x \mid v^{\pm} \ket$.
In general 
\be
\bra v^{\pm} \mid \sigma_i^x \sigma_j^x \mid v^{\pm} \ket =
 (-\i)^{j-i} \bra v^{\pm} \mid \tau^-_i \prod_{k<j} \tau_k^+\tau_k^-  \tau_j^+ \mid v^{\pm} \ket 
\quad .
\label{korrx}
\ee
Of course $\i$ denotes $\sqrt{-1}$.
Using eq.(\ref{kev}) the $\tau^{\pm}_j$  can be expressed in
 terms of ladder operators $a_k$ and $b_k$, i.e.
\be
\tau^{\mu}_j=\sum_{k=0}^{L+1} (\phi^-_k)^{\mu}_j a_k + (\phi^+_k)^{\mu}_j b_k \quad .
\label{tauina}
\ee
Because $(\phi_0^{\pm})_j^{\mu}$ is only different from zero if either
$j=0$ and $\mu=+$ or $j=L+1$ and $\mu=-$ (compare eq.(\ref{zeroladderop}) and eqs.(\ref{a_dagger}),(\ref{a}))
 we have
\be
\bra v^{\pm} \mid \tau^-_i \prod_{k<j} \tau_k^+\tau_k^-  \tau_j^+ \mid v^{\pm} \ket=
\bra vac \mid \tau^-_i \prod_{k<j} \tau_k^+\tau_k^-  \tau_j^+ \mid vac \ket \quad .
\label{vpmvac}
\ee
For simplicity,  we will denote $\bra vac| \hat{O}|vac \ket$ by 
$\bra \hat{O} \ket$ in the following 
 where $\hat{O}$ denotes an arbitrary operator.
Using Wick's theorem we are left with the calculation of the Pfaffian of the antisymmetric Matrix $\bf A$, i.e.
\be
\bra \tau^-_i \prod_{k<j} \tau_k^+\tau_k^-  \tau_j^+ \ket = \mbox{Pf} {\bf A} 
\quad,
\label{PfA}
\ee
where  
\be
\fl
{\bf A} = \left(
\begin{array}{cccccc}
0 & \bra \tau_i^- \tau_{i+1}^+ \ket & \bra \tau_i^- \tau_{i+1}^- \ket & \bra \tau_i^- \tau_{i+2}^+ \ket & \cdots & \bra \tau_i^- \tau_j^+ \ket\\
\bra \tau_{i+1}^+ \tau_i^- \ket & 0 & \bra \tau_{i+1}^+ \tau_{i+1}^- \ket & \bra \tau_{i+1}^+ \tau_{i+2}^+ \ket & \cdots & \bra \tau_{i+1}^+ \tau_j^+ \ket\\
\bra \tau_{i+1}^- \tau_i^- \ket & \bra \tau_{i+1}^- \tau_{i+1}^+ \ket & 0 & \bra \tau_{i+1}^- \tau_{i+2}^+ \ket & \cdots & \bra \tau_{i+1}^- \tau_j^+ \ket\\
\bra \tau_{i+2}^+ \tau_i^- \ket & \bra \tau_{i+2}^+ \tau_{i+1}^+ \ket & \bra \tau_{i+2}^+ \tau_{i+1}^- \ket & 0 & \cdots & \bra \tau_{i+2}^+ \tau_j^+ \ket \\
\vdots & \vdots & \vdots & \vdots & \ddots & \vdots \\
\bra \tau_j^+ \tau_i^- \ket & \bra \tau_j^+ \tau_{i+1}^+ \ket & \bra \tau_j^+ \tau_{i+1}^- \ket & \bra \tau_j^+ \tau_{i+2}^+ \ket & \cdots & 0 \\
\end{array}
\right)
\quad .
\label{matrixA}
\ee
We want to remind the reader that the Pfaffian of a $2n \times 2n$ 
antisymmetric matrix $A$ with matrix elements 
$a_{ij}$ is defined by
\be
\mbox{Pf} A = \frac{1}{n!2^n}\sum_{\sigma\in S_{2n}}\mbox{sgn}(\sigma)
a_{\sigma(1)\sigma(2)} a_{\sigma(3)\sigma(4)} \cdots a_{\sigma(2n-1)\sigma(2n)} \quad ,
\ee
where $S_{2n}$ denotes the symmetric group of degree $2n$. 

The expectation values of the basic contractions of pairs which form
the entries of ${\bf A}$
 are evaluated using eq.(\ref{tauina}). Due to the property of $\bra vac |$ 
(cf.\  eq.(\ref{leftvac}))
we obtain 
\be
\bra \tau_i^{\mu} \tau_j^{\nu} \ket = \sum_{k=0}^{L+1} (\phi_k^-)_i^{\mu}(\phi_k^+)_j^{\nu} \quad .
\label{basiccontra}
\ee
In general,  no further simplification is possible.

Nevertheless there exist two special cases where
 the calculation of the Pfaffian can be reduced
to the calculation of a determinant. Namely, this is possible if no diagonal boundary terms
are present or if $\alpha_-=\alpha_+$ and $\beta_-=\beta_+$.
This reduction uses the 
additional relations for the eigenvectors given by eqs.(\ref{crucial}) or (\ref{crucial2})
respectively. Note that in this paper the general relation 
$(\mbox{Pf}{\bf A})^2=\det{\bf A}$ is of no use 
because 
we are exactly interested in the sign of $\eta$.

In the absence of diagonal boundaries we can 
use  eq.(\ref{crucial}) to simplify eq.(\ref{basiccontra}). 
Using similar arguments as LSM we also obtain  a determinant
representation for  the correlation functions, however, the contributing 
contractions are different
from theirs. In fact, using eq.(\ref{crucial}) results in
\be
\bra \tau_i^{\mu} \tau_j^{\nu} \ket = 0 \quad \mbox{for $i+j$ even}  
\ee
and thus 
 the correlation functions are given by subdeterminants of the 
$(L+1) \times (L+1)$ matrix
\be
{\bf D} = \left( 
\begin{array}{cccc}
\bra \tau_0^- \tau_{1}^+ \ket & \bra \tau_0^- \tau_1^- \ket
 & \bra \tau_0^- \tau_3^+ \ket & \cdots \\
\bra \tau_2^+ \tau_1^+ \ket & \bra \tau_2^+ \tau_1^- \ket
 & \bra \tau_2^+ \tau_3^+ \ket & \cdots \\
\bra \tau_2^- \tau_1^+ \ket & \bra \tau_2^- \tau_1^- \ket 
& \bra \tau_2^- \tau_3^+ \ket & \cdots \\
\vdots & \vdots & \vdots & \ddots
\end{array}
\right)
\quad .
\label{bfD}
\ee
Denoting by ${\bf D}^i_j$ the matrix ${\bf D}$ after elimination of the first 
$i$ rows and columns and the last $L+1-j$ rows and columns we can write
\be
\mbox{Pf}{\bf A} = \i^{j-i}f_{ij}\det {\bf D}^i_j \quad ,
\label{detD}
\ee
where
\be
f_{ij}=\left\{
\begin{array}{cc}
-\i & \mbox{if $i$ even and $j$ odd} \\
\i  & \mbox{if $i$ odd and $j$ even} \\
1   & \mbox{otherwise} 
\end{array}
\right.
\quad .
\ee
The calculation of Pf$\bf A$ also simplifies if 
$\alpha_+=\alpha_-$ and $\beta_+=\beta_-$. In this case we can utilize
 eq.(\ref{crucial2}) to obtain
\be
\bra \tau_i^{\mu} \tau_j^{\mu} \ket = 0 \quad .
\ee
This again results in a determinant representation of Pf$\bf A$, i.e.
\be
\mbox{Pf}{\bf A}=\left|
\begin{array}{cccc}
\bra \tau_i^- \tau_{i+1}^+ \ket & \bra \tau_i^- \tau_{i+2}^+ \ket &
\cdots  &\bra \tau_i^-\tau_j^+ \ket \\
\bra \tau_{i+1}^- \tau_{i+1}^+ \ket & \bra \tau_{i+1}^- \tau_{i+2}^+ \ket & 
\cdots & \bra \tau_{i+1}^-\tau_j^+ \ket \\
\vdots & \vdots &  & \vdots \\
\bra \tau_{j-1}^- \tau_{i+1}^+ \ket & \bra \tau_{j-1}^- \tau_{i+2}^+ \ket & 
\cdots &  \bra \tau_{j-1}^-\tau_j^+ \ket \\
\end{array}
\right|
\quad .
\label{detG}
\ee
Thus in eq.(\ref{PfA}) we are left with the 
calculation of the  subdeterminants of a $(L+1) \times (L+1)$ matrix $\bf G$ with elements
\be
G_{kl}=\bra \tau_{k-1}^-\tau_l^+ \ket \quad ,
\ee
where $1 \leq k \leq L+1$ and $1 \leq l \leq L+1$.

By simple modifications one can
generalize the results of this section to excited states of the form
(\ref{even}) or (\ref{odd}). 
The argument runs as follows: Any state given by elementary excitations
can be regarded as the vacuum state $|vac'\ket$ of a new set of 
ladder operators, where the $a_k$ and $b_k$ of the excited fermions
are interchanged.
 This corresponds to an interchange of the eigenvectors 
 $\phi_k^-$ and $\phi_k^+$ (see eqs.(\ref{a}) and (\ref{a_dagger})). 
Thus the calculation of correlation functions for the states
in (\ref{even}) and (\ref{odd}) is equivalent to the calculation
of the correlation functions for the states $|{v^+}'\ket$ defined 
in the same way as $|v^+\ket$ in
eq.(\ref{vplusminus}) but now using 
\be
|vac'\ket = \prod_{j=1}^r b_{k_j}|vac\ket \quad ,
\ee
as vacuum state
where $r$ is even or odd respectively and $k\neq 0$. 
The left vacuum defined by eq.(\ref{leftvac}) has to be modified analogously.
As consequence,  we only have to replace eq.(\ref{basiccontra}) by
\be
\bra \tau_i^{\mu} \tau_j^{\nu} \ket = 
\sum_{k \mbox{\tiny\ unexc.}} (\phi_k^-)_i^{\mu}(\phi_k^+)_j^{\nu} +
\sum_{k \mbox{\tiny\ exc.}} (\phi_k^+)_i^{\mu}(\phi_k^-)_j^{\nu} \quad .
\ee
\end{section}

\begin{section}{Calculation of $\bra\sigma_{L+1}^x\ket$ }

If the diagonal boundary terms are absent or if 
$\alpha_+=\alpha_-$ and $\beta_+=\beta_-$, 
we can make use of eqs.(\ref{detD}) or (\ref{detG}) respectively 
 in order to calculate 
\be
\eta=\bra v^+| \sigma^x_{L+1}| v^+ \ket \quad .
\ee
Recall that we need the value of $\eta$
to decide whether the $(++)$-sector of the space of states of $H_{long}$ 
is given by the states of the form (\ref{even}) or (\ref{odd}).

Thus if no diagonal boundary terms are present, we have to calculate 
$\det {\bf D}$ where ${\bf D}$ is defined by eq.(\ref{bfD}).
This determinant can be written as the product of two
determinants 
\be
(-1)^L\det {\bf D}   = \det {\bf A}_g \det {\bf A}_u  
\label{DA}
\ee 
where
\be
\fl
{\bf A}_g=\left(
\begin{array}{cccc}
(\phi^-_1)_0^- & (\phi^-_2)_0^- & (\phi^-_3)_0^- & \cdots \\
(\phi^-_1)_2^- & (\phi^-_2)_2^- & (\phi^-_3)_2^- & \cdots \\
(\phi^-_1)_2^+ & (\phi^-_2)_2^+ & (\phi^-_3)_2^+ & \cdots \\
\vdots & \vdots & \vdots & \ddots 
\end{array}
\right)
{\bf A}_u=\left(
\begin{array}{cccc}
(\phi^+_1)_1^- & (\phi^+_1)_1^+ & (\phi^+_1)_3^- & \cdots \\
(\phi^+_2)_1^- & (\phi^+_2)_1^+ & (\phi^+_2)_3^- & \cdots \\
(\phi^+_3)_1^- & (\phi^+_3)_1^+ & (\phi^+_3)_3^- & \cdots \\
\vdots & \vdots & \vdots & \ddots 
\end{array}
\right)
\quad .
\ee     
The factor $(-1)^L$ is due to permutations of rows and columns which
can be seen by comparing the product ${\bf A}_g{\bf A}_u$ with ${\bf D}$.   
Without loss of generality we may assume that the eigenvectors are
normalized in such a way that they satisfy eq.(\ref{nodiagn}) with $c=1$.
Using eq.(\ref{crucial}) we then find
\be
{\bf A}_g {\bf A}_g^t={\bf 1} \quad , \quad {\bf A}_u {\bf A}_u^t=-{\bf 1}      
\quad .
\label{au2ag2}
\ee
Due to the special form of the matrix M 
(see eq.(\ref{form})) and to eq.(\ref{nodiagn})
the two matrices ${\bf A}_g$ and ${\bf A}_u$ are related
by a matrix ${\bf M}_{g\to u}$ via
\be
{\bf \Lambda}^{-1}{\bf M}_{g\to u}{\bf A}_g={\bf A}_u^t \quad ,
\label{AA}
\ee
where ${\bf \Lambda}_{kk'}= \Lambda_k \delta_{kk'}$ with $k > 0$ .
The matrix ${\bf M}_{g\to u}$ is given by
\be
{\bf M}_{g\to u}=\left(
\begin{array}{ccccc}
-G'^t & L & 0 & 0 & 0 \\
0 & -L^t & \ddots & 0 & 0 \\
0 & 0 & \ddots & L & 0 \\
0 & 0 & 0 & -L^t & K' \\
\end{array}
\right)
\ee
for $L$ odd or 
\be
{\bf M}_{g\to u}=\left(
\begin{array}{ccccc}
-G'^t & L & 0 & 0 & 0 \\
0 & -L^t & \ddots & 0 & 0 \\
0 & 0 & \ddots & L & 0 \\
0 & 0 & 0 & -L^t & L \\
0 & 0 & 0 &  0 & -K'^t \\
\end{array}
\right)
\ee
for $L$ even, where we denoted
by $G'$ the matrix $G$ of eq.(\ref{matrices})
with the second row eliminated
and by $K'$ the matrix $K$ of eq.(\ref{matrices})
with the first column eliminated.
Note that the eliminated rows and columns contain
only entries which are equal to zero.

Using the eqs.(\ref{AA}) and (\ref{au2ag2}) in eq.(\ref{DA}), we obtain
\be
\det{\bf D} = (-1)^L\det{\bf \Lambda}^{-1} \det{\bf M}_{g\to u} \quad .
\ee
The value of $\det{\bf M}_{g\to u}$ can be computed in an elementary way
\be
\det{\bf M}_{g\to u}=\left\{
\begin{array}{cc}
 -(\alpha_-\beta_+ + \alpha_+\beta_-)/4^{L+1} & \mbox{for L odd} \\
 -\i(\alpha_-\beta_+ + \alpha_+\beta_-)/4^{L+1} & \mbox{for L even}
\end{array}
\right.
\ee
Plugging this into eq.(\ref{detD}), we end up with
\be
\mbox{Pf}{\bf A} = (-\i)^{L+1} \frac{\alpha_-\beta_++\alpha_+\beta_-}{4^{L+1}\prod_{k\neq 0} \Lambda_k}        
\quad .
\label{endup}
\ee
If $\alpha_-=\alpha_+$ and $\beta_-=\beta_+$, 
 we obtain the same result by using equation (\ref{detG}) 
and performing a similar calculation. 
Combining eqs.(\ref{fjintau}),(\ref{vpmvac}) and (\ref{PfA}) 
with eq.(\ref{endup}) we are left in both cases with
\be
\label{final}
\eta = \bra v^+| \sigma^x_{L+1}| v^+ \ket =
(-1)^{L+1} \frac{\alpha_-\beta_++\alpha_+\beta_-}{4^{L+1}\prod_{k\neq 0} \Lambda_k} 
\quad .
\label{etaresult}
\ee
Notice that in these cases the expression of $\eta$ does not depend on
$\alpha_z$ and $\beta_z$.
It is not possible to calculate the product of all eigenvalues 
in eq.(\ref{etaresult}) in general, but
the squared product of eigenvalues can be calculated from $\det M'$ where $M'$ denotes 
 the matrix $M$ with the second 
and the last but one row and column eliminated. In both cases this yields
\be
\det{\bf \Lambda}^2=(-1)^{L+1}\det M' =
(\alpha_-\beta_+ + \alpha_+\beta_-)^2/4^{2L+2} \quad .
\ee
If the Hamiltonian is hermitian, the product of eigenvalues is simply given by
$\prod_{k\neq 0} \Lambda_k = |\alpha_-\beta_++\alpha_+\beta_-|/4^{L+1}$.
This also holds for
Hamiltonians with only real entries because in these cases the product is  
real and positive. Thus, we can apply  equation (\ref{final}) directly.
Otherwise one would have to know all the eigenvalues explicitly.

Note that the value of $\eta$ 
 can only change by variation of the boundaries if one crosses a point in
the parameter space at which an additional
mode with Re$\Lambda_k=0$ and $k\neq 0$ exists.
This is due to the fact that the eigenvalues
of $H$ and $H_{long}$ are continuous functions of the boundary parameters.
But at a point satisfying the above condition 
the value of $\eta$ is not well defined 
because the corresponding ladder operators $a_k$ and $b_k$ are
not well defined as already mentioned in section 2.1. Therefore
one can aquire a change of sign in $\eta$ by passing through such a point.

If $H$ is hermitian the condition to have a mode with
 Re$\Lambda_k=0$ and $k\neq 0$ is equivalent 
to the existence of an additional zero mode.
The presence of such a zero mode corresponds to the root $z=-1$ in the 
polynomial given by eq.(\ref{pol}). This implies the condition 
 $\alpha_-\beta_++\alpha_+\beta_-=0$. 
Thus if $H$ is hermitian, i.e. $\alpha_+=\alpha_-^*, 
\beta_+=\beta_-^*$,
we have 
to only distinguish the two regions Re$(\alpha_-\beta_+)>0$ and
Re$(\alpha_-\beta_+)<0$. 
Thus, we conclude that if $H$ is hermitian we obtain the following
expression for $\eta$
\be
\eta = (-1)^{L+1} \mbox{sign}(\re(\alpha_-\beta_+)) \quad .
\label{etahermitesch}
\ee
The results of this section, namely eqs.(\ref{final}) and (\ref{etahermitesch}),
 allow us to calculate the ground state energy 
of H for the exactly solvable cases of table I. This will be the subject of section 12.
\end{section}

\section{Ground state energies for the Hamiltonian $H$ in the exactly 
solvable cases}

In section 9 we have shown that the ground state of $H$ corresponds
either to $|v^+ \rangle$ or to $b_{lowest} |v^{-}\rangle$ where 
$b_{lowest}$ is the creation operator 
corresponding to the fermion 
energy with the smallest real part which we will denote by
$2 \Lambda_{lowest}$ in the following, and the $|v^{\pm}\rangle$
are defined in eq. (\ref{vplusminus}). Which of these two 
states corresponds to  the ground state
depends on the eigenvalue $\eta$ of $| v^{+} \rangle$ with respect to the 
operation of $\sigma^{x}_{L+1}$. The eigenvalue $\eta$ is either
$+1$ or $-1$. If
$\eta = 1$, then the ground state corresponds to
$|v^+ \rangle$, and the ground state energy
of $H$ is equal to the ground state energy of $H_{long}$.
If $\eta = -1$, then the ground state corresponds to
$b_{lowest} |v^{-}\rangle$
and the ground state energy of $H$ is given by the sum
of the ground state energy of $H_{long}$ and $2 \Lambda_{lowest}$. 
For the exactly
solvable cases, the ground state energies for $H_{long}$ are already contained
in table II.

If at least one of the following conditions is satisfied:
\begin{itemize}
\item[a)] $H_{long}$ is hermitian
\item[b)] $H_{long}$ has no $\sigma^{z}$ boundary terms ($\az = 0 = \bz$)
\item[c)] $\a = \g$ and $\b = \d$,
\end{itemize}
the value of $\eta$ can be easily calculated by using the explicit
formulas of section 11. In the cases b) and c),
the expression for $\eta$ is given in eq. (\ref{final}) in terms of 
$\a,\b,\d$ and
$\g$ and the eigenvalues $\Lambda_k$ of $H_{long}$. In the case a)
where $H_{long}$ is hermitian, $\eta$ is given by eq.\ (\ref{etahermitesch})
in terms of $\a$ and $\b$ alone.
In the other cases, one would have to calculate the Pfaffian of the 
matrix $\bf A$
given by eq. (\ref{matrixA}) using different methods than the ones we used
in the sections 10 and 11 to decide which of the states
$|v^+ \rangle$ or $b_l |v^{-}\rangle$ corresponds to
the ground state of $H$. 

To determine the ground state energy
for a given Hamiltonian $H$ of the type $a) -c)$
one has to calculate the expression for $\eta$ given by eq.\ (\ref{final})
or eq.\ (\ref{etahermitesch}). 
Analytically, the ground state energy for $H$
can be calculated with our methods
only for the cases given in table I where
the polynomial factorizes into cyclotomic polynomials and where one knows
the whole spectrum of $H_{long}$. These cases are given in terms of
the parameters $A,B,C,D$ and $E$. In order to put our machinery 
to work, we need the 
corresponding parameters $\a,\g,\d,\b,\az$ and $\bz$.
Since the transformation
from $A,B,C,D$ and $E$ to $\a,\g,\d,\b,\az$ and $\bz$
(which is given in eqs. (\ref{rates})) is non--linear
and leads from 5 to 6 variables, the choice of the parameters
$\a,\g,\d,\b,\az$ and $\bz$ for a given set $A,B,C,D$ and $E$ is not unique,
creating some freedom of choice.

For all the cases listed in table I, we solved  the eqs. (\ref{rates})
for $\a,\g,\d,\b,\az$ and $\bz$ and allowed only 
solutions which additionally satisfy
one of the conditions a)-c) above. These solutions are listed in table III.
The choices of boundary parameters obtained from the ones given in 
table III by application of an obvious
similarity transformation to the Hamiltonian $H$
such as reflecting the Hamiltonian in the middle
of the chain or applying a transformation of the form given by
eq.\ (\ref{trafo}) to the Hamiltonian
are not explicitly listed. 

Table III should be understood as follows:
Let us first comment on the choice of the signs in the cases
where we give two alternative signs for the boundary parameters. 
The signs in the third and in the fourth column can always be
chosen independently of each other.
However, in the upper half of the table,
the signs of $\a$ and $\g$ cannot be chosen independently
(see e.\ g.\ case 2 or 4) whereas in the lower part of the table,
they can be chosen independently (e.\ g.\ in the cases 9 even
and 9 odd as indicated by the condition  $\g \d = \pm 1$).

Now we turn to the one--parameter families 
depending on the parameter $s$. Here, 
the conditions a)-c) often lead to restrictions
for the value of $s$ which are indicated in the fifth column. 
The choice of $\a,\g,\d$ and $\b$ given in the fourth column
determines $\a \b +\g \d$. In most of the cases from table I
the product of all values of $\Lambda_k$ is positive
(or zero), and consequently
the sign of $\a \b +\g \d$ multiplied by $(-1)^L$ yields
the sign of $\eta$. If $\eta < 0$, the ground state energy of $H$
is given by $E_0(H_{long}) + 2 \Lambda_{lowest}$ where $E_0(H_{long})$ can be
taken from table II.
If $\eta > 0$, the ground state energy of $H$ is given by
$E_0(H) = E_0(H_{long})$. However, there are some cases where
eigenvalues with vanishing real part but non--vanishing imaginary part
appear in the spectrum of $H_{long}$. This happens
for example in case 10 for negative values of $s$ and may in general
happen in the cases 9, 10, 14 and 15. In 
these cases, the sign of the product
of all eigenvalues is not uniquely defined.  
Here, it is impossible to decide
which of the two vectors $|v^+ \rangle$ and $b_{lowest} |v^{-}\rangle$
corresponds to the ground state of $H$.

In some of the cases,
we always find $\a \b +\g \d=0$. Here, an additional zero mode appears
in the spectrum of $H_{long}$ as already mentioned at the end
of section 3. Therefore the 
energies of $|v^+ \rangle$ and $b_n |v^{-}\rangle$
where $b_n$ is the creation operator for the additional fermion
with energy zero are the same and the ground state energy of $H$ is again
given by $E_0(H_{long})$. This is also indicated in table III in the last 
column.

\begin{table}
\caption{Exactly solvable cases from table I. Details needed for the projection method}
\begin{center}
\lineup
\footnotesize
\begin{tabular}{@{}l|l|ll|l|l|l}
\br
case &   $L$ & $\az$ & $\bz$ & $\a$\;\;\;\;\;\;  $\g$\;\;\;\;\;\;  $\b$ \;\;\;\;\;\; $\d$& $s$
& $2 \Lambda_{lowest}$ \\ \mr
\\
1.) & arb. &0 & 0 & $\a \g =1\;\;\; \b \d =\frac{1}{2}\;\;\; \g \d = 
\pm \frac{(1 \pm \i)}{2}$& & $ \sin{\frac{\pi}{4L+6}}$\\
2.) & arb. &0 & 0 & $\a \g =1\;\;\; \b = \d =0\;\;\; 
$& & 0  \\
 & arb. & 0 & $\pm \frac{1}{\sqrt{2}}$ & $\a =\pm \frac{1}{\sqrt{2}} 
e^{\i \phi}$
  \;\;\; $\g = \pm \frac{1}{\sqrt{2}} e^{-\i \phi}$ \;\;\;  
$\b = \d =0$ & & 0 \\
3.) & arb. &0 & 0 & $\a \g =\frac{1}{2}\;\;\; \b = \d =0\;\;\;
$& & 0 \\
4.) & arb. &0 & $\pm \frac{1}{\sqrt{2}}$ & $\a =\pm e^{\i \phi} \;\;\;  \g =
\pm e^{-\i \phi} \;\;\;\b = \d =0$ & &0  \\
 & arb. & 0 & $\pm \frac{\i}{\sqrt{2}}$ & $\a = \g =0\;\;\; \b = \d =\sqrt{2}$ 
& & 0\\
5.) & even &0 & 0 & $\a \g =\frac{1}{2}\;\;\; \b \d =\frac{1}{2}\;\;\; \g \d =
\pm \frac{1}{2}$& & $ \sin{\frac{\pi}{2L+6}}$ \\
5.) & odd &0 & 0 & $\a \g =\frac{1}{2}\;\;\; \b \d =\frac{1}{2}\;\;\; \g \d =
\pm \frac{\i}{2}$& &0 \\
6.) & even &0 & 0 & $\a \g =\frac{1}{2}\;\;\; \b \d =\frac{1}{2}\;\;\; \g \d =
\pm \frac{\i}{2}$& & 0\\
6.) & odd &0 & 0 & $\a \g =\frac{1}{2}\;\;\; \b \d =\frac{1}{2}\;\;\; \g \d =
\pm \frac{1}{2}$& & $ \sin{\frac{\pi}{2L+4}}$ \\
7.) & even &0 & 0 & $\a \g =\frac{1}{2}\;\;\; \b \d =1\;\;\; \g \d =
\pm \frac{1}{\sqrt{2}} $& & $ \sin{\frac{\pi}{2L+4}}$ \\
7.) & odd &0 & 0 & $\a \g =\frac{1}{2}\;\;\; \b \d =1\;\;\; \g \d =
\pm \frac{\i}{\sqrt{2}} $& &0 \\
8.) & even &0 & 0 & $\a \g =\frac{1}{2}\;\;\; \b \d =1\;\;\; \g \d =
\pm \frac{\i}{\sqrt{2}} $& & 0\\
8.) & odd &0 & 0 & $\a \g =\frac{1}{2}\;\;\; \b \d =1\;\;\; \g \d =
\pm \frac{1}{\sqrt{2}} $& & $ \sin{\frac{\pi}{2L+4}}$ \\ \mr
9.) & even &0 & 0 &$\a \g =1\;\;\; \b \d =1\;\;\; \g \d =
\pm \sqrt{s}$& & min($ \sin{(\frac{\pi}{2L+2}\pm \frac{\i \ln{s}}
{2L+2}}))$\\
 & even &0 & 0 &$\a \g =1\;\;\; \b \d =1\;\;\; \g \d =
\pm \frac{1}{\sqrt{s}}$& & min($ \sin{(\frac{\pi}{2L+2}\pm 
\frac{\i \ln{s}} {2L+2}}))$\\
 & even &0 & $\pm \frac{\i}{\sqrt{2}}$ & $\a = \g = \pm \frac{1}{\sqrt{2}}
\;\;\; \b=\d = \pm \sqrt{2} \;\;\; \g \d = \pm 1 $& $s = 1$ & 
$ \sin{\frac{\pi}{2L+2}}$ \\
9.) & odd &0 & 0 &$\a \g =1\;\;\; \b \d =1\;\;\; \g \d =
\pm \i \sqrt{s}$& & min($ \sin{(\frac{\pi}{L+1}\pm\frac{\i \ln{s}} 
{2L+2}}))$\\
& odd &0 & 0 &$\a \g =1\;\;\; \b \d =1\;\;\; \g \d =
\pm \frac{\i}{\sqrt{s}}$& & min($\sin{(\frac{\pi}{L+1}\pm
\frac{\i \ln{s}} {2L+ 2}}))$\\
 &odd &0 & $\pm \frac{\i}{\sqrt{2}}$ & $\a = \g = \pm \frac{1}{\sqrt{2}}
\;\;\; \b=\d = \pm \sqrt{2} \;\;\; \g \d = \pm 1 $& $s = -1$ & 
$ \sin{\frac{\pi}{2L+2}}$\\
10.) & even &0 & 0 &$\a \g =\frac{s+1}{2} \;\;\; \b \d =\frac{s+1}{2 s}
\;\;\; \g \d = \pm \frac{(s^{1/2} + s^{-1/2})}{2}$& & min($ 
\sin{\frac{\pi}{2L+2}}, \frac{1}{2} (s^{1/2} + s^{-1/2}))$\\
 &even   &0 & $\pm \frac{\i}{\sqrt{2}}$ & $\a= \g = \pm \frac{1}{\sqrt{2}}
\;\;\; \b = \d = \pm \sqrt{\frac{2+s+1/s}{2}}$& & min($ 
\sin{\frac{\pi}{2L+2}}, \frac{1}{2} (s^{1/2} + s^{-1/2}))$\\
10.) & odd &0 & 0 &$\a \g =\frac{s+1}{2} \;\;\; \b \d =\frac{s+1}{2 s}
\;\;\; \g \d = \pm \i \frac{(s^{1/2} + s^{-1/2}}{2})$& & 0\\
 & odd &0 & $\pm \frac{\i}{\sqrt{2}}$ &$\a= \g = \pm \frac{1}{\sqrt{2}}
\;\;\; \b = \d = 0$& $s =-1$ & 0\\
11.) &arb. &0 & $\pm \frac{1}{\sqrt{2}}$& $\a = \g = \b = \d =0$ & $s = -1$ &
0\\
12.) &arb. & $\pm \frac{1}{\sqrt{2}}$& $\pm \frac{1}{\sqrt{2}}$& 
$\a = \g = \b = \d =0$ & $s = 1$ & 0\\
 &arb. & $\mp \frac{\i}{\sqrt{2}}$& $\pm \frac{\i}{\sqrt{2}}$& 
$\a = \g = 0\;\;\; \b = \d =
\pm \sqrt{2}$ & $s = 1$ & 0 \\
13.) &arb. & $\mp \frac{\i}{\sqrt{2}}$& $\pm \frac{\i}{\sqrt{2}}$& 
$\a = \g = 0\;\;\; \b = \d =
1 $ & $s = -1$ & 0\\
14.) & even & $\pm\frac{\i}{\sqrt{2s}}$ & $ \pm \i \sqrt{\frac{s}{2}}$ &
$\a = \g =\sqrt{\frac{1+s}{s}} \;\;\; \b = \d = \sqrt{1+s}$ & & 
min($ \sin{\frac{\pi}{2L}},\frac{1}{2} (s^{1/2} + s^{-1/2}))$\\
14.) & odd& $\pm\frac{1}{\sqrt{2}}$ & $\mp \frac{1}{\sqrt{2}}$ &
$\a = \g = \b = \d =0$ & $ s= -1$ & 0\\
15.) & even & $\pm\frac{1}{\sqrt{2}}$ & $\pm\frac{1}{\sqrt{2}}$ &
$\a = \g = \b = \d =0$ & $ s= -1$ & 0 \\
15.) & odd& $\mp \frac{\i}{\sqrt{2s}}$ & $ \pm \i \sqrt{\frac{s}{2}}$ &
$\a = \g =\sqrt{\frac{1+s}{s}} \;\;\; \b = \d = \sqrt{1+s}$ & & 
min($ \sin{\frac{\pi}{2L}},\frac{1}{2} (s^{1/2} + s^{-1/2}))$\\
16.) &arb. & $\pm\frac{1}{\sqrt{2s}}$ & $ \pm \sqrt{\frac{s}{2}}$ &
$\a = \g = \b = \d =0$ & & 0\\ 
\mr \br
\end{tabular}
\end{center}
\end{table}

\section{Guide}

In this article, we explained how to diagonalize
the XX--quantum spin chain of length $L$ with diagonal and non--diagonal
boundary terms 
defined in eq.\ (\ref{Ham}). Here we give a resumee
of our method which the reader may use as a guide on how to use our
results. This guide should be seen as a user--friendly cooking recipe.
It has two parts, the first deals with the spectrum, the 
second with the eigenvectors. As one will notice, the guide does
not follow the sections in a chronological way.

In order to find the eigenvalues and --vectors of $H$ we
start by considering a different Hamiltonian 
$H_{long}$ which is obtained from $H$ by 
appending two additional sites $0$ and $L+1$ (see eq.\ (\ref{H_long}))
so that the expression for $H_{long}$ is 
bilinear in Majorana (Clifford) operators, see eq. (\ref{hI}). 
$H_{long}$ can be diagonalized in terms of free fermions,
fixing the representation we are working in. The spectrum and
the eigenvectors of $H$ in the Fock representation
can be retrieved from the ones found for
$H_{long}$ by a projection method described below. 

\subsection{Eigenvalues of $H_{long}$}
The diagonalization of $H_{long}$ is described
in section 2. The spectrum is given in terms of  $L+2$ single
fermionic energies $2 \Lambda_n$ (see eq.\ (\ref{spectrum})). The values of
$\Lambda_n$ can be obtained from
a  $(2L+4) \times (2L+4)$
matrix M (eq.\ (\ref{matrix_M})). Since $M = - M^{t}$, 
the $2L+4$ eigenvalues of this matrix
appear in pairs $\pm \Lambda_n$. The
necessary $L+2$ eigenvalues are taken by convention as the values
with positive real part. As explained in the text, zero is always an
eigenvalue of $M$. This corresponds to a  fermionic zero mode.
As long as we consider $H_{long}$, the zero mode
$\Lambda_0 =0$ appears in the spectrum and the ground state is at least twofold
degenerate. As we are going to see,
the zero mode does not appear in the spectrum of $H$, therefore we are going to 
call it spurious zero mode. 
However, the eigenvectors of $M$ corresponding to the spurious zero mode
will be needed in the derivation of the eigenvectors of $H$.
   
The values of $\Lambda_n$ can be expressed using eq.\ (\ref{Lambda}) 
in terms of the zeros 
of a complex polynomial of degree $2L+4$ (see eq.\ (\ref{pol})). 
To find the zeros of the polynomial analytically, we have looked in 
a systematic way for factorizations 
of the polynomial into cyclotomic polynomials.
We determined all possible factorizations  up to five factors and
found some examples for factorizations in six factors.
These results are listed in table I (the 
parameters $A, B, C, D$ and $E$ appearing in table I 
are defined by eq.\ (\ref{rates}) in terms of the boundary parameters
of the Hamiltonian). 
For the cases where we did not find any factorizations of the polynomial,
the zeros of the polynomial and therewith the fermionic energies
can still be calculated numerically. Since the polynomial has degree $2L+4$,
this is much easier than a straightforward
numerical diagonalization of the Hamiltonian which has dimension $2^{L}
\times 2^{L}$.

By studying the solutions of the polynomial equation (\ref{pol}),
we find special $L$--independent solutions in some cases.  
They correspond to boundary bound states as will be shown in \cite{paper2}.
 
The ground state energy of $H_{long}$ (which is by convention the energy with
the smallest real part)
is obtained in eq.\ (\ref{spectrum}) by
subtracting the Fermi sea. 
In table II we listed
the corresponding expressions for
the ground state energies of $H_{long}$ (which are at least twofold degenerate)
for the cases 
where the polynomial factorizes
into cyclotomic polynomials. Some properties of the ground state energies will
be discussed in section 14.

In section 6, we give the expressions for the spectrum of $M$ 
in some of the "exactly solvable" cases. A list of 
the ground state energies of $H_{long}$ for all "exactly solvable" cases
can be found in section 7. Section 8 contains the spectrum of $M$
and the ground state energy of one example
of a Hamiltonian with asymmetric bulk terms which can be treated with the
results developed in this article by using the similarity transformation
between the Hamiltonian given by eq.\ (\ref{Ham}) and the one given by
eq.\ (\ref{Ham_nh}). 
This transformation changes the boundary parameters 
according to  eq.\ (\ref{trans}).

\subsection{Eigenvalues and ground state energy of $H$} 
Finding the eigenvalues of the original Hamiltonian $H$ is more involved.
As shown in section 9, to find the spectrum $H$ we have to look at an even or
an odd number of fermionic excitations with respect to the 
lowest energy of $H_{long}$.  We
disregard the spurious zero mode in the calculation of the number of the
fermionic excitations.
Whether one has an even or an odd number of fermionic excitations in the
spectrum of $H$ depends on the value
of a parameter $\eta$ defined by eq. (\ref{eta})
which is either $+1$ or $-1$ (see section 9 for details). The way it
is computed is going to be explained below.
If $\eta= +1$, the spectrum of
$H$ consists of an even number of fermionic excitations with respect
to the ground state energy of $H_{long}$ and the ground state energy
of $H$ is the same as the one of $H_{long}$. If $\eta = -1$, the eigenvalues of
$H$ are given by an odd number of fermionic excitations and the ground state
energy of $H$ is the sum of the ground state energy of $H_{long}$ and
the fermionic energy with the smallest real part which we call
$2 \Lambda_{lowest}$. 

If on top of the spurious zero mode
another fermionic excitation is zero, 
the ground state energy of $H$ is non--degenerate and 
identical to the ground state energy of $H_{long}$ and the spectrum of $H$
is given by all even and odd combinations of fermionic excitations.
If a second fermionic excitation is zero, the whole spectrum of $M$ is 
twofold degenerate.
So in these cases one does not need to calculate the value of $\eta$.

At this point we restrict our discussion to the 
cases, where we have derived explicit formulas for the 
parameter $\eta$: 
\begin{itemize}
\item[a)] $H_{long}$ (and therewith  $H$) is hermitian 
\item[b)] $H_{long}$ has no $\sigma^{z}$ boundary terms ($\az = 0 = \bz$)
\item[c)] $\a = \g$ and $\b = \d$.
\end{itemize}
For the other cases this guide is not sufficient since they are 
much more complicated and we  have not obtained
simple formulas for the parameter $\eta$.

In case a), $\eta$
is given by eq.\ (\ref{etahermitesch})). Notice that only two 
boundary parameters appear in the expression for $\eta$. 
In the cases b) and c), the expression
for  $\eta$ is given by eq.\ (\ref{final}) in terms of the parameters
of the non--diagonal boundary terms and the eigenvalues of M. 

\subsubsection{Analytical results for the ground state energy of $H$}
\hspace{3cm}

\noindent
If the Hamiltonian 
additionally belongs to one of the "factorizable" cases, 
the ground state energy of $H$ can be calculated analytically.
These cases are listed in table III. To calculate the ground state energy
of $H$ for a particular choice of boundary parameters given in the third
and fourth column of table III, one has to proceed as follows:
First one checks the value of $2 \Lambda_{lowest}$ given in the last column.
If $2 \Lambda_{lowest}= 0$, the ground state energy of $H$ is identical to
that of $H_{long}$ (which is listed in table II). In the cases 
where $2 \Lambda_{lowest} \neq 0$, one has to know the  value of $\eta$ 
to obtain the ground state energy of $H$. This value 
is obtained by using
formula (\ref{final}) in the cases b) and c) (the eigenvalues $\lambda_n$
are given by the zeros of the factorized polynomials listed in table I
via eq.\ (\ref{Lambda}))
and formula (\ref{etahermitesch}) in the case a).

The fermionic energy with the smallest real part $2 \Lambda_{lowest}$
which has to be added
to the ground state energy of $H_{long}$ if $\eta = -1$ is listed in the
last column of table III.
If $\eta = 1$ the ground state energy of $H$ can be taken directly
from table II.

Many of the exactly solvable cases depend on an arbitrary free parameter $s$
(see table I and III). In table III, these 
 $s$--dependent 
cases can be separated into two categories. For the cases 11 -- 13
and for some choices of the parameters in the cases 9, 10, 14 and 15,
the conditions a)--c) fix the parameter $s$  to 
some  particular value which can be found in column 5 of table III.
For the cases 9, 10, 14, 15 and  16 there are also possible choices 
of the boundary parameters where this is not the case.
In the examples 9, 10, 14 and 15 
it may happen that one cannot make a definite statement about
the value of $\eta$, if $s$ is
chosen in such a way that the $s$--dependent eigenvalue of $M$
has a vanishing real part, but a non--vanishing imaginary part.
The reason lies in the fact
that our convention to choose the fermionic energies as the ones with positive
real part becomes ambiguous in this case.

\subsubsection{Numerical calculation of the ground state energy of $H$}
\hspace{4cm}

\noindent
Even if the Hamiltonian one is interested in does not belong to one of the
factorizable cases, but fulfils conditions a), b) or c),
one can still use the formulas (\ref{etahermitesch}) and (\ref{final})
to decide what the ground state of $H$ is. If $H$ is hermitian, 
$\eta$ can be read off directly from eq.\ (\ref{etahermitesch}), 
in the cases b)
and c) one additionally needs the spectrum of $M$ to compute 
the value of $\eta$ (see eq.\ (\ref{final})). The eigenvalues of
$M$ can be calculated numerically by solving the polynomial equation
(\ref{pol}) or by diagonalizing $M$ numerically. Inserting them 
into eq.\ (\ref{spectrum}) yields the
ground state energy of $H_{long}$.

\subsection{Eigenvectors of $H_{long}$, $H$ and $M$}

Up to now we have described how to find the eigenvalues and the ground state 
energies for $H_{long}$ and for $H$. 
Let us now turn to the eigenvectors.

The eigenvectors of $H_{long}$
are given in a fermionic Fock representation (compare eq. \ref{spectrum}).

The eigenvectors of $H$ are
given in the same Fock representation, however they all lie either in the
even or the odd part of the Fock space where we again to not count
the spurious zero mode. 
If the value of $\eta$
is $+1$, the ground state of $H$ corresponds to $| v^{+} \ket$
(see eq.\ (\ref{vplusminus})) and all excited states are of the form
given in  eq.\ (\ref{even}).
If $\eta = -1$, the ground state
of $H$ corresponds to  $b_{lowest} | v^{-} \ket$ where $| v^{-} \ket$
is also defined in eq.\ (\ref{vplusminus}) and $b_{lowest}$
is the creation operator corresponding to the fermion
energy with the smallest real part. For the exactly solvable cases
it can be read off table III. The excited states are
described by eq.\ (\ref{odd}).

In section 10, we describe how to calculate expectation values of
$\sigma$--operators. For this calculation, 
one can either transform the expression for
the eigenstates of $H$ in the spin representation or alternatively,
one can transform the expression for the $\sigma$--operators into
the fermionic (Fock) representation. We have  chosen the second 
possibility. The transformation from the $\sigma$--operators to
the fermionic operators $a_k$ and $b_k$ is given in
eqs.\ (\ref{tau}) and (\ref{tauina}) where the $(\phi^{\pm}_k)^{\mu}_j$
are the components of the eigenvectors of $M$ defined by 
eq.\ (\ref{eigenvalue_problem}) 
(where we use the  notation fixed by eq.\ (\ref{eigenvector})).  
Thus, to use this transformation one needs to know the eigenvectors 
of $M$. We will now describe how to find them analytically in the cases
where the zeros of the polynomial are known,
following the method described in section 3. One first solves 
eq.\ (\ref{m}) to express 
$\vp_1$ as function of $\vpd_1$ 
where the coefficients $\Omega_{ij}$ with $i,j=1,2$
are given by eqs. (\ref{omega11}) -- (\ref{omega22}). 
If $x \neq \i$, the solution for
$\vp_1$ is inserted in eqs. (\ref{b})--(\ref{g}), and the
results for the coefficients $a,b,f$ and $g$ used in eq.\
(\ref{solution_bulk}) for $x \neq 1$ respectively
eq.\ (\ref{lim}) for $x=1$ yield expressions for $\vp_j$ and $\vpd_j$. 
The entries of the eigenvector
$\phi^{\pm}$ 
are then given by 
eq.\ (\ref{varphi2}) in terms of $\vp_j$ and $\vpd_j$. 
In this notation, they still depend on the variable
$x$. 

The values of $x$ are obtained as solutions of
the polynomial equation (\ref{pol}). The polynomial is given in the
variable $z = x^2$. The eigenvectors for the eigenvalues $\Lambda_n$
with positive
real part and $x_n \neq  \i$
are obtained by choosing a square root $x_n = \sqrt{z_n}$
for each zero $z_n \neq \pm1$ of the polynomial (such 
that the real part of $x_n$
is positive). 
Observe that due to the quadratic
relation between $\Lambda_n$ and $x_n$ (eq.\ (\ref{x_lambda}))
the values  $x_n$ and $1/x_n$ lead to the same eigenvalue and
to the same eigenvector. 

For the eigenvectors corresponding to
the  eigenvalues 
$\Lambda_n$ with negative real part one takes $x_n = -\sqrt{z_n}$.
The last free parameter $\vpd_1$ is fixed by the normalization
conditions
given by eq.\ (\ref{ev2}). Equations (\ref{ev2}) and (\ref{ev0}) are
equivalent to the anticommutation relations for the fermionic operators.

For $x_n = \i$ the equations (\ref{bulkeqd})--(\ref{rightb2})
have to be solved using the ansatz given by
eq.\ (\ref{solution_bulk}) for $\vp_j$ and $\vpd_j$. 
Details of this calculation as well as a derivation of the conditions 
for the appearance of additional zero modes on top of the
spurious zero mode in the spectrum of $H_{long}$
can be found in the appendix.

\subsection{One-- and two--point functions of the $\sigma^{x}$--operators}

If no $\sigma^{z}$--boundary terms are present in the Hamiltonian or if the
condition $\a = \g$ and $\b = \d$ is met, we obtained formulas for the
one--and two--point functions of the $\sigma_j^x$--operator 
for both chains $H$ and $H_{long}$.
For $H_{long}$, we considered the ground states given by the eigenstates
of $\sigma_0^x$ and $\sigma_{L+1}^x$. Remember that $H_{long}$ has a twofold
degenerate ground state due to the spurious zero mode. In the fermionic 
language, this corresponds to a vacuum, $|vac \rangle$ and an excited zero
mode, $|0 \rangle$. Due to the symmetry that $H_{long}$ commutes with
$\sigma_0^x$ and $\sigma_{L+1}^x$ we can pick out the two ground states
$|v^{\pm}\rangle = |vac \rangle \pm |0 \rangle$ (see eq.\ (\ref{vplusminus}))
as eigenstates of
$\sigma_0^x$ and $\sigma_{L+1}^x$. As far as $H$ is concerned, 
its ground state is given either by $|v^{+} \rangle$ or by $b_{lowest}
|v^{-}\rangle $ where $b_{lowest}$ is the creation 
operator of the fermion corresponding to the energy with the 
smallest real part as explained above. 

The one--point functions of the $\sigma_j^x$--operator
are non--trivial due to the presence of the non--diagonal boundary 
terms. Without them, they would be zero which is a well--known fact from the
XX--chain.

The one-- and two--point functions are up to a factor Pfaffians 
(see eqs.\ (\ref{fjintau}), (\ref{korrx}) and (\ref{PfA})) 
of the matrix  {\bf A}
given by eq.\ (\ref{matrixA}).
If no $\sigma^z$--boundary terms are present,  we can further reduce this
expression to a determinant 
as given by eq.\ (\ref{detD}).  
If $\a = \g$ and $\b = \d$,  the determinant is given by
eq.\ (\ref{detG}). These simplifications are possible because 
some of the so--called basic contractions of pairs
of the form given by eq.\ (\ref{basiccontra}) vanish due to
the relations (\ref{crucial}) and (\ref{crucial2}) obtained in section 2. 
To determine the ground state of $H$, we use the results of this calculation
to determine  the 
magnetization $\bra v^+| \sigma_{L+1}^x| v^+ \ket$
in section 11 to find the value of the parameter $\eta$ as 
given by eq.\ (\ref{final}).

Our calculation with slight modifications
also applies to expectation values of the $\sigma^x_j$--operator
with respect to excited states. This is explained in the last paragraph 
of section 10.

\subsection{List of the results which are going to be used in the following two
articles}

Here we give a list of results that we will use in the following two
articles:

\begin{itemize}
\item Second article \cite{paper2}\\
In order to calculate expectation values of the $\sigma^z_j$--operator
and the $\sigma^x_j$--operator for arbitrary position $j$ and lattice 
length $L$, we need 
\begin{enumerate} 
\item the transformation from the
$\sigma$--operators to the fermionic operators 
(eqs.\ (\ref{tau}) and (\ref{tauina}))
\item the eigenvectors of $M$ (see section 3)
and the eigenvalues $\Lambda_n$ of $M$ (eqs.\ (\ref{Lambda}) and (\ref{pol}))
\item the expressions for the eigenstates of $H$ in the Fock representation
(eqs.\ (\ref{even}), (\ref{odd}) and (\ref{vplusminus})) and the value of
the parameter $\eta$ defined by eq.\ (\ref{eta})
\item the formulas for the one-- and two--point correlation 
functions of $\sigma_j^x$ derived in section 10
\end{enumerate}
\item Third article \cite{paper3}
\begin{enumerate}
\item
For the calculation of the excitation spectrum of $H$ in the limit of large $L$
we need the polynomial equation (eq.\ (\ref{pol})) and the projection mechanism
(see section 9).
\item
For the expressions of the ground state energies in the exactly solvable cases
in the limit of large $L$ we need the results of table II and III.
\item
For the construction of the magnetic charge operator, we need the
eigenvectors of $M$ (and therewith the eigenvalues, see above).
\end{enumerate}
\end{itemize}

\section{Discussion}
\subsection{Observations on the expressions for the ground state energies}

Up to now we have described how to find the eigenvalues, the ground state
energies and the eigenvectors for $H_{long}$ (eq.\ (\ref{H_long})) and for $H$
(eq.\ (\ref{Ham})). 
We now turn to the discussion of
the results of our analytical calculations for the cases
where the polynomial (eq. (\ref{pol}))
can be factorized into cyclotomic polynomials. 
The expressions of the ground state energies of $H_{long}$ and of $H$ 
are given in terms of trigonometric functions only (see tables II and III).
It is remarkable that they appear 
in spite of non--diagonal boundary terms in the Hamiltonians.
This reflects the integrability of the model. 

Furthermore, notice that  for the one--parameter families of exact solutions 
corresponding to  the
cases 10 -- 16 from table II (where the free parameter is called $s$) 
the ground state energy has 
an $L$--independent (but $s$--dependent) term which appears additively to the
$L$--dependent part of the ground state energy. We will show in \cite{paper2}
that these $L$--independent contributions to the ground state energy
of $H$  are related to boundary bound states.

In some special situations which only appear for non--hermitian Hamiltonians
the expressions for the ground state energy of $H$
exhibit a rather peculiar behaviour 
with respect to variations of $L$. Namely, considering one of the cases 
10, 14 and 15 and choosing the  parameter $s$ in such a way that 
$\eta = -1$ one observes that for $L$ less than a limiting length
$L_{limit}$ (which depends on $s$) 
the fermionic energy 2 $\Lambda_{lowest}$ (which has to be added
to the ground state energy of $H_{long}$ to obtain the ground state energy
of $H$) is
given by twice the  $L$--independent expression that appears
in the ground state energy of $H_{long}$, but
with opposite sign (see table III). Therefore this term appears with different
sign in the expression for the ground state energy of $H$ than in the
expression for the ground state energy of $H_{long}$.
However,
if $L$ is larger than $L_{limit}$, a level crossing in the fermionic
spectrum appears and
another $L$--dependent fermionic energy becomes smaller than the
$L$--independent energy from before. Then this $L$--dependent fermionic energy
has to be added to the ground state of $H_{long}$ 
instead of the $L$--independent
term from before, and the 
$L$--independent part  no longer switches its sign when going from the 
ground state energy of $H_{long}$ to the one of $H$.  

Now we discuss the degeneracies in the spectrum of $H$. Degeneracies
may appear due to
doubly degenerate fermionic energies. 
In the cases where the polynomial can be factorized
into cyclotomic polynomials, twofold degenerate fermionic energies can be
identified by quadratic factors appearing in the factorized form of the
polynomial (see table I). -- Notice that this observation
does not apply to the quadratic
factors of the form $(1-z)^2$ since the polynomial $p(z)$ given by
eq. (\ref{pol}) has to be  divided by this term. --
In the case where the spectrum of $H$
consists of an odd number of fermionic excitations, twofold 
degenerate fermionic energies  also lead to
a twofold degenerate ground state. 
The degeneracies in the spectrum of 
$H$ are also reflected in the partition functions in \cite{paper3}. 

\subsection{Open questions}

Some questions could not be clarified within the framework of this
article. 
\begin{itemize}
\item[a)]
It is not clear whether table I from section 5 shows all
possible cases where the polynomial factorizes
in six or more factors given by cyclotomic polynomials.
Perhaps it is also possible to find different factorizations
of the polynomial for other boundary parameters
which also allow to compute all zeros
analytically.

\item[b)]
From table III one sees that two Hamiltonians may have different
boundary terms and still have the same spectrum
(given by the zeros of the same polynomial) 
e.\ g.\  in the case 2 where the Hamiltonian $H$ with $\a \g =1$ and
$\b=\d=\az=\bz=0$ has the same spectrum as the Hamiltonian with
boundary parameters $\bz = \frac{1}{\sqrt{2}}, \az =0,
\a = \frac{1}{\sqrt{2}} e^{\i \phi},\g =  \frac{1}{\sqrt{2}} e^{-\i \phi},
\b = \d =0$.
This fact gives rise to unknown similarity transformations
which remain to be made explicit.

\item[c)]
For the exactly solvable 
one--parameter families  10, 14 and 15 with free parameter $s$
we observed a surprising behaviour of the expression of
the ground state energy of $H$ in the case where the parameter $\eta = -1$. 
Namely, by increasing the lattice length 
$L$ and reaching a certain value $L_{limit}$ which is given in terms of
$s$, the $L$--independent contribution to the 
ground state energy suddenly switches its sign as explained above.
The physical origin of this  phenomenon is not clear.

\end{itemize}

\vspace{1cm}

\section*{Acknowledgements}

We would like to thank V. Rittenberg for many discussions and 
constant encouragement.
We are grateful to A. Nersesyan for valuable suggestions
and fruitful discussions. We would also like to acknowledge F.C. Alcaraz,
A.A. Belavin, F.H.L. Essler, G. Harder, R.M. Kaufmann, K. Krebs, 
L. Mezincescu and D. Zagier for helpful discussions.

\appendix
\section{Appearance of fermionic zero modes in the spectrum of $H_{long}$}
In this appendix, we show how to find the  eigenvectors of the matrix 
$M$ corresponding to the eigenvalue zero. This procedure will provide
conditions for the boundary parameters which are equivalent to the existence 
of additional zero modes on top
of the spurious zero mode. 
These conditions are already mentioned without proof in section 3.

One might guess that the conditions on the boundary parameters we obtain by 
constructing the eigenvectors are already contained in the polynomial given 
by eq.(\ref{pol}). We will see that this is indeed the case, if we only
consider hermitian boundary terms. In the general, non--hermitian case
this is not true. The polynomial might have more zeros corresponding to
eigenvalues $\lambda=0$ of the matrix $M$ than eigenvectors can be
constructed. Therefore, in this case, $M$ is not diagonalizable. 

We are first going to deal with
 the explicit construction of the eigenvectors. Afterwards we will consider
the polynomial equation (\ref{pol}). 

\subsection{Construction of eigenvectors}
According to eq.(\ref{x_lambda}) $\lambda=0$ corresponds to $x=\pm\i$.
So we solve the boundary equations (\ref{leftb1})-(\ref{rightb2}) using 
the solution of the bulk equations (\ref{bulkeqd}) given by (\ref{bulks})
with $x=\i$, i.e.
\be
\vp_j=a\i^j+b\i^{-j}\qquad , \qquad \vpd_j=g(-\i)^j + f(-\i)^{-j} \quad ,
\label{bulkszero}
\ee
where $0<j<L+1$.
This can now be used to rewrite the boundary equations in terms of $a,b,g,f$ and
$\vp_0,\vpd_0,\vp_{L+1},\vpd_{L+1}$. Introducing the new parameters
\be
r_{\alpha}^{\pm}=\left( \frac{1}{\sqrt{2}} \pm\i\alpha_z \right) \quad , \quad
r_{\beta}^{\pm} =\left( \frac{1}{\sqrt{2}} \pm\i\beta_z \right)  
\ee
we obtain from the left boundary
\ba
\vp_0=\vpd_0 \label{leftb1t} \\
\alpha_-(a-b)=\alpha_+(f-g) 
 \label{leftb10} \\
\rap a +
\ram b -
\alpha_+\vp_0 =  0 \label{leftb20a} \\  \phantom{}
\ram g +
\rap f -
\alpha_-\vp_0 =  0                  
\quad . \label{leftb20b}
\ea
The equations from the right boundary give
\ba
(-1)^L \rbm a -
\rbp b +
\i^{L+1}\beta_+\vp_{L+1} =  0 \label{rightb10a} \\  \phantom{}
\rbp g -
(-1)^L \rbm f +
\i^{L+1}\beta_-\vp_{L+1} =  0               
\label{rightb10b}
 \\
\beta_+(g+(-1)^Lf)=-\beta_-(b+(-1)^La) \label{rightb20} \\
\vp_{L+1}=-\vpd_{L+1} \label{rightb2t} \quad .
\ea
Since $\vpd_0$ and $\vpd_{L+1}$ appear only in the eqs.(\ref{leftb1t}) 
and (\ref{rightb2t}),
we have to solve the homogeneous system of 6 linear equations given by
 eqs.(\ref{leftb10})-(\ref{rightb20}) for the 6 unknowns $a,b,f,g,\vp_0,\vp_{L+1}$.
The vector components $\vpd_0$ and $\vpd_{L+1}$ can then be directly read off from
eq.(\ref{leftb1t}) respectively eq.(\ref{rightb2t}). To have non--trivial solutions
for the $6 \times 6$ 
system of equations (\ref{leftb10})-(\ref{rightb20}) the determinant 
of the corresponding $6 \times 6$ matrix has to vanish. This is equivalent to a 
condition on the boundary parameters $\alpha_+,\alpha_-,\beta_+,\beta_-$, i.e.
\be
\alpha_+\beta_-+\alpha_-\beta_+=0 \label{cond0} \quad .
\ee
At this point, it is not obvious how many solutions of the 
eqs.(\ref{leftb10})-(\ref{rightb20}) 
may exist. Thus we are going to solve them explicitly.
In order to do this, we will treat eqs.(\ref{leftb20a})-(\ref{rightb10b}) and
eqs.(\ref{leftb10}) and (\ref{rightb20}) separately. 
We first solve eqs.(\ref{leftb20a})-(\ref{rightb10b}) for $a,b,f,g$ and then
check for consistency with eqs.(\ref{leftb10}) and (\ref{rightb20}).

Eqs.(\ref{leftb20a}) and (\ref{rightb10a})  and eqs.(\ref{leftb20b}) and (\ref{rightb10b})
 can be rewritten as
\be
\fl
R_{ab} \left(\begin{array}{l} a \\ b \end{array} \right )
= \left(\begin{array}{c} \alpha_+\vp_0 \\
	              -\i^{L+1}\beta_+\vp_{L+1} 
\end{array}\right) \ , \
R_{gf} \left(\begin{array}{l} g \\ f \end{array} \right )
= \left(\begin{array}{c} \alpha_-\vp_0 \\
		      - \i^{L+1}\beta_-\vp_{L+1}
\end{array}\right) \ ,
\label{Reqs}
\ee
where the $2\times2$-matrices $R_{ab}$ and $R_{gf}$ are given by
\be
R_{ab}= \left( \begin{array}{cl}
\rap & \ram \\
(-1)^L \rbm & -\rbp
\end{array}\right) \ , \ 
R_{gf}= \left( \begin{array}{lc}
\ram & \rap \\
\rbp & (-1)^{L+1} \rbm
\end{array}\right) \ .
\label{Rab}
\ee
The determinants of $R_{ab}$ and $R_{gf}$ have the same values and are given in terms
of $\alpha_z,\beta_z$ by
\be
\det R_{ab}=\det R_{gf} = \left\{ \begin{array}{ll}
2\beta_z\alpha_z-1 & \mbox{for $L$ even} \\
-\sqrt{2}\i(\alpha_z+\beta_z) & \mbox{for $L$ odd} 
\end{array}\right. \quad . 
\label{det0}
\ee
Once we know the general solution of eq.(\ref{Reqs}),
we only have to verify which 
of them simultaneously solve the eqs.(\ref{leftb10}) and (\ref{rightb20}).
Solving eq.(\ref{Reqs}) one has to distinguish two cases:
(1) $\det R_{ab}\neq 0$ , (2) $\det R_{ab}=0$ .
Let us first deal with case (1).

If $\det R_{ab}\neq 0$ the matrices $R_{ab}$ and $R_{gf}$ can be inverted in order to
 solve eq.(\ref{Reqs}).
Doing this we obtain 
\be
a=\frac{-1}{\det R_{ab}}\left(
\alpha_+ \rbp \vp_0 -
\i^{L+1}\beta_+ \ram \vp_{L+1}
\right) \ , \label{a0}
\ee
\be
b=\frac{-1}{\det R_{ab}}\left( (-1)^L
\alpha_+ \rbm \vp_0 +
\i^{L+1}\beta_+ \rap \vp_{L+1}
\right) \ ,
\ee
\be
g=\frac{-1}{\det R_{ab}}\left( (-1)^L
\alpha_- \rbm \vp_0 -
\i^{L+1}\beta_- \rap \vp_{L+1}
\right) \ ,
\ee
\be
f=\frac{-1}{\det R_{ab}}\left( 
\alpha_- \rbp \vp_0 + 
\i^{L+1}\beta_- \ram \vp_{L+1}
\right) \ . \label{b0}
\ee
Substituting this into eq.(\ref{leftb10}) gives 
\be
(\alpha_+\beta_- + \alpha_-\beta_+)\vp_{L+1} = 0 \quad ,
\ee
whereas eq.(\ref{rightb20}) leads to
\be
(\alpha_+\beta_- + \alpha_-\beta_+)\vp_0 = 0 \quad .
\ee
Thus, if eq.(\ref{cond0}) is satisfied and if $\det R_{ab}\neq 0$,
we obtain two eigenvectors of $M$ corresponding to the eigenvalue
$\lambda=0$ on top of the spurious zero mode
because $\vp_0$ and $\vp_{L+1}$ can be chosen independently of each other. 

Let us now turn to case (2), i.e. $\det R_{ab}=0$.
Because this condition gives different conditions on the 
non-diagonal boundary terms for $L$ even and $L$ odd
respectively (see eq.(\ref{det0})) we discuss
these cases separately.

We will first turn to the case where $L$ is odd. 
Here  we have $\alpha_z=-\beta_z$ which can be read off from
eq.(\ref{det0}).
Using this in eq.(\ref{Rab}) we can rewrite eq.(\ref{Reqs}) 
as follows: 
\be
\alpha_+\vp_0=\i^{L+1}\beta_+\vp_{L+1} \quad , \quad \alpha_-\vp_0=-\i^{L+1}\beta_-\vp_{L+1} 
\quad , \label{0l+1o}
\ee
\be
\alpha_+\vp_0=\rbm a+\rbp b \quad , \quad \alpha_-\vp_0= \rbm f+ \rbp g 
\quad . \label{abfgo}
\ee
For $L$ even we have according to eq.(\ref{det0}) $\alpha_z=\frac{1}{2\beta_z}$.
Using this equality, eq.(\ref{Reqs}) reads:
\be
\sqrt{2}\beta_z\alpha_+\vp_0=\i^L\beta_+\vp_{L+1} \quad , \quad
\sqrt{2}\beta_z\alpha_-\vp_0=-\i^L\beta_-\vp_{L+1} \quad ,
\label{0l+1e}
\ee
\be
\i\sqrt{2}\beta_z\alpha_+\vp_0=\rbp b - \rbm a \quad , \quad
\i\sqrt{2}\beta_z\alpha_-\vp_0=\rbp g - \rbm f \quad .
\label {abgfe}
\ee
Note that at least one of the parameters $\rbm,\rbp$ is different from zero. 
Thus the eqs.(\ref{abfgo}) and (\ref{abgfe}) can be solved either
for $a$ and $f$ or for $b$ and $g$
respectively.
Note also that due to eqs.(\ref{0l+1o}) and (\ref{0l+1e}) $\vp_0$ and $\vp_{L+1}$ are no longer
independent of each other, 
if one of the parameters $\alpha_+,\alpha_-,\beta_+,\beta_-$
is different from zero in contrast to the case (1). 

If all of the parameters $\alpha_+,\alpha_-,\beta_+,\beta_-$ are vanishing
the eq.(\ref{0l+1o}) respectively eq.(\ref{0l+1e})  are satisfied 
automatically. The same holds for eqs.(\ref{leftb10}) and (\ref{rightb20}).
Thus we simply have to solve eqs.(\ref{abfgo}) and (\ref{abgfe}) yielding
\be
a=(-1)^L\frac{\rbp}{\rbm}b \quad , \quad
f=(-1)^L\frac{\rbp}{\rbm}g \quad \mbox{for $\rbm\neq 0$}
\ee
\be
b=(-1)^L\frac{\rbm}{\rbp}a \quad , \quad
g=(-1)^L\frac{\rbm}{\rbp}f \quad \mbox{for $\rbp\neq 0$}
\quad .
\ee
Since on the one hand the parameters $a$ and $f$ or $b$ and $g$ respectively and on
the other hand the two vector components $\vp_0,\vp_{L+1}$ can be chosen independently
we obtain a set of four eigenvectors corresponding to the eigenvalue
 $\lambda=0$ on
top of the spurious zero mode.

If one of the parameters  $\alpha_+,\alpha_-,\beta_+,\beta_-$ is different from zero 
we may solve eqs.(\ref{abfgo}) and (\ref{abgfe}) for $a$ and $f$ or $b$ and $g$
respectively and use the result in eq.(\ref{leftb10}) to obtain 
\be
\alpha_-b=\alpha_+g \quad \mbox{for $\rbm\neq 0$}
\label{lbv}
\ee 
\be
\alpha_-a=\alpha_+f \quad \mbox{for $\rbp\neq 0$}
\quad .
\ee 
Additionally using eq.(\ref{0l+1o}) respectively eq.(\ref{0l+1e}), 
we obtain from
eq.(\ref{rightb20})
\be
\beta_+g=-\beta_-b \quad \mbox{for $\rbm\neq 0$}
\label{rbv}
\ee
\be
\beta_+f=-\beta_-a \quad \mbox{for $\rbp\neq 0$}
\quad .
\label{rbw}
\ee
Due to the condition (\ref{cond0}), it is always possible to solve eq.(\ref{lbv})-(\ref{rbw})
and eq.(\ref{0l+1o}) respectively eq.(\ref{0l+1e}) 
by leaving two variables  undetermined. 
The remaining four unknowns can then be given in terms
of these two. This allows the construction of two further eigenvectors
corresponding to the  eigenvalue $\lambda=0$.

For instance, let us assume $\alpha_+\neq 0,\ \rbm\neq 0$ and $L$ odd.
Eq.(\ref{0l+1o}) is then solved by $\vp_0=\i^{L+1}\frac{\beta_+}{\alpha_+}\vp_{L+1}$
whereas eqs.(\ref{lbv}) and (\ref{rbv}) are solved by $g=\frac{\alpha_-}{\alpha_+}b$.
The parameters $a$ and $f$ are then fixed by eq.(\ref{abfgo}). Thus $a,f,g,\vp_0$ 
are given in terms of $b$ and $\vp_{L+1}$ which can be chosen independently. 

Let us briefly summarize the results of this section. We looked for eigenvectors
corresponding to the eigenvalue zero. At this point we want to remind the reader
that there always exist  at least two eigenvectors corresponding to 
the eigenvalue zero, namely the ones which 
are related to the spurious zero mode.
We found exactly two further eigenvectors corresponding to the eigenvalue zero of $M$ if 
and only if one of the two following conditions is satisfied :
\begin{enumerate}
\item
$\alpha_+\beta_- + \beta_+\alpha_- = 0$ and $\beta_z \neq \frac{1}{2\alpha_z}$ for $L$ even       
respectively $\alpha_z\neq-\beta_z$ for $L$ odd.
\item
$\alpha_+\beta_- + \beta_+\alpha_- = 0$ and at least one of the parameters $\alpha_{\pm},\beta_{\pm}$
is different from zero.
\end{enumerate}
There exist four further eigenvectors corresponding to the eigenvalue zero of $M$ if and
only if 
\begin{enumerate}
\item
$\alpha_{\pm}=\beta_{\pm}=0$ and $\beta_z = \frac{1}{2\alpha_z}$ for $L$ even 
respectively $\alpha_z=-\beta_z$ for $L$ odd.
\end{enumerate}
There are no other possibilities to have further eigenvectors corresponding to 
the  eigenvalue 
zero.

\subsection{Zeros of the polynomial at $z=-1$}

In this subsection we consider the polynomial equation (\ref{pol})
in order to check whether the matrix $M$ may have more eigenvalues $\lambda=0$
than eigenvectors can be constructed. Since in general $M$ is non--hermitian,
this may 
indeed be the case. For hermitian $M$, i.e. hermitian boundary conditions,
we will recover the conditions on the boundary terms which
we already obtained in the previous 
section.

Since the polynomial equation is given in terms of the variable $z$
which is related to the eigenvalues via eq.(\ref{Lambda}), 
an additional eigenvalue $\lambda=0$ on top of the spurious zero mode
corresponds to a root of the polynomial
at $z=-1$. 
The necessary condition to have at least one eigenvalue $\lambda=0$ 
on top of the spurious zero mode therefore translates to
\ba
& q(-1)=0 \nonumber \\
\Leftrightarrow & D=1+A+B+2 C  \label{D=}\\
\Leftrightarrow & \alpha_-\beta_+ + \alpha_+\beta_- = 0 \label{polcond0}
\quad ,
\ea
where the parameters $A,B,C,D,E$ are defined by eqs.(\ref{rates}).
These zeros will always appear in pairs
since with $z$ also $\frac{1}{z}$ is a zero of $q(z)$.
In order to find a condition for the existence of a root at $z=-1$
with higher multiplicity than two, we
have to consider the second derivative of $q(z)$. Using eq.(\ref{D=}) we obtain
for even $L$
\ba
& \partial_z^2 p(z)_{|z=-1}=0  \nonumber \\
\Leftrightarrow & 2(E-1)^2 + (3+2A+B+2C) L  -C L^2  =0 \label{polcond1e} \quad ,
\ea
whereas for odd $L$ we get
\be
1-C-B+4E + (3+2A+B+2C) L  -C L^2  =0  \label{polcond1o} \quad .
\ee
Let us now consider eqs.(\ref{polcond0})-(\ref{polcond1o}) for the case of
hermitian boundaries.
Then eq.(\ref{polcond0}) implies that at least one of the parameters 
$\alpha_+=\alpha_-^*$ or $\beta_+=\beta_-^*$
is equal to zero. Without loss of generality we may assume that $\alpha_+=0$.
This implies immediately $C=0$ and $3+2A+B=2|\beta_+|^2(1+2\alpha_z^2)$.
Using eq.(\ref{polcond1e}) we obtain  for $L$ even
\be
2|\beta_+|^2(1+2\alpha_z^2)=-\frac{2(2\alpha_z\beta_z-1)^2}{L} \quad .
\ee
Since this equality can only be valid if the RHS and the LHS vanish
simultaneously,
 we conclude that in the hermitian case further zeros at $z=-1$ only exist 
if
\be
\alpha_+=\beta_+=0 \quad , \quad \alpha_z=\frac{1}{2\beta_z} \quad .
\label{condhe}
\ee
{From} eq.(\ref{polcond1o}) we get for $L$ odd
\be
2|\beta_+|^2(1+2\alpha_z^2)=-\frac{4(\beta_z+\alpha_z)^2}{L} \quad ,
\ee
which can only be satisfied if
\be
\alpha_+=\beta_+=0 \quad , \quad \alpha_z=-\beta_z \quad .
\label{condho}
\ee
Further conditions for the existence of more than four zeros at $z=-1$ can be
 derived in the same manner as eq.(\ref{polcond1e}) and eq.(\ref{polcond1o})
respectively. Solving eqs.(\ref{polcond1e}) and (\ref{polcond1o}) for $A$ and
calculating the fourth derivative of $p(z)$ at $z=-1$ gives the conditions
\be
\fl
4-8 E+ E^2+ 12 L (1-E^2) + L^2 (2 E^2-3B-4C+8E+5)+L^4 C =0
\label{polcond2e}
\ee
for $L$ even and
\be
\fl
C+8E+B+6E^2+5+12 L (1-E^2) +L^2(7-2C-8E-B+6E^2) + L^4 C =0
\label{polcond2o}
\ee
for $L$ odd respectively.
Using eqs.(\ref{condhe}) in eq.(\ref{polcond2e}) gives
\be
\left(\sqrt{2}\beta_z+\frac{1}{\sqrt{2}\beta_z}\right)^2=0
\label{nixe}
\quad ,
\ee
whereas substitution of eqs.(\ref{condho}) in eq.(\ref{polcond2o})
yields
\be
-2\beta_z^2=\frac{1+L}{L-1} \quad .
\label{nixo}
\ee
Since neither eq.(\ref{nixe}) nor eq.(\ref{nixo}) can be satisfied
by any $\beta_z\in \BRT$ we conclude that in the hermitian case
we have at most four zeros at $z=-1$.
It is no surprise that for hermitian boundaries 
the conditions on the boundary parameters obtained in this subsection are equivalent to the ones 
of the previous subsection. 
However, if $M$ is non--hermitian, the conditions derived in this 
subsection have more solutions than
the ones of the previous section. 
Therefore it may happen that the polynomial has more zeros corresponding to an 
eigenvalue $\lambda=0$ than eigenvectors can be constructed.
This implies that $M$ is non--diagonalizable 
for certain choices of boundary terms.

\listoftables

\newpage
\section*{References}
\begin{thebibliography}{99}
\bibitem{LSM}{Lieb E., Schultz T. and Mattis D. 1961 {\em Ann. Phys.}
{\bf 16} 407}
\bibitem{McCoy}{McCoy B. 1968 {\em Phys. Rev.} {\bf 173} 531}
\bibitem{Barouch_McCoy} {Barouch E. and McCoy B. 1971 {\em Phys. Rev. A} {\bf 3}
 786}
\bibitem{MBA}{McCoy B., Barouch E. and Abraham B. 1971
{\em Phys. Rev. A} {\bf 4} 2331}
\bibitem{Vaida_Tracy}{Vaidya H.  and Tracy C. 1978 {\em Physica A} {\bf 92} 1}
\bibitem{Bariev}{Bariev R.Z. 1978 {\em Phys. Lett} {\bf 68A} 175}
\bibitem{Vladimir_Haye} {Hinrichsen H. and Rittenberg V. 1993
{\em Phys. Lett. B} {\bf 304} 115}
\bibitem{toy_paper} {Bilstein U. and Wehefritz B. 1997 \JPA {\bf 30} 4925}
\bibitem{Guinea} {Guinea F. 1985 {\em Phys. Rev. B} {\bf 32} 7518;\\
Guinea F., Hakim V. and Muramatsu A. 1985 {\em Phys. Rev. Lett} {\bf 54} 263}
\bibitem{non-hermitian}{Hatano N. and Nelson D.R. 1996
{\em Phys. Rev. Lett} {\bf 77} 570;\\
Efetov K.B. 1997 {\em Phys. Rev. Lett} {\bf 79} 491; 
1997 {\em Phys. Rev. B} {\bf 56} 9630;\\
Janik R.A., Nowak M.A., Papp G. and Zahed I. {\em Localization 
Transition from Free Random Variables}, preprint cond-mat/9705098;\\
Mudry C., Simons B.D. and Altland A. 1998 {\em Phys. Rev. Lett} {\bf 80} 4257;\\
Fukui T. and Kawakami N. {\em Breakdown of the Mott insulator: Exact 
solution of an asymmetric Hubbard model}, preprint cond-mat/9806023}
\bibitem{Fabian_Valdimir}{Essler F.H.L. and Rittenberg V. 1996 \JPA {\bf 29}
3375}
\bibitem{Vega}{de Vega H.J. and Gonzales-Ruiz A. 1994 \JPA {\bf 27} 6129}
\bibitem{Inami}{Inami T. and Konno H. 1994 \JPA {\bf 27} L913}
\bibitem{Bariev_Peschel}{Bariev R.Z. and Peschel I. 1991 {\em Phys. Lett. A}
{\bf 153} 166\\
Hinrichsen H., Krebs K. and Peschel I. 1996 \ZP {\bf B100} 105}
\bibitem{Luca}{Mezincescu L. and Nepomechie R.I. {\em Fractional--Spin Integrals of Motion for the Boundary Sine--Gordon Model at the Free Fermion Point},
preprint hep-th/9709078}
\bibitem{Ameduri}{Ameduri M., Konik R. and LeClair A. 1995 \PL {\bf B354} 376}
\bibitem{Kane_Fisher}{Kane C.L. and Fisher M.P.A. 1992 {\em Phys. Rev. B}
{\bf 46} 15233}
\bibitem{field_theory} {Ghoshal S. and Zamolodchikov A.B. 1994 {\em Int.
J. Mod. Phys. A} {\bf 9} 3841;\\
Fendley P. and Saleur H. 1994 {\em Nucl. Phys. B} {\bf 428} 681;\\ 
Ghoshal S. 1994 {\em Int. J. Mod. Phys. A} {\bf 9} 4801}
\bibitem{Skorik_Saleur}{Skorik S. and Saleur H. 1995 \JPA {\bf 28} 6605}
\bibitem{Jordan_Wigner} {Jordan P. and Wigner E. 1928 {\em Z. f. Phys.} {\bf 47}
631} 
\bibitem{Haye} {Hinrichsen H. 1994 \JPA {\bf 27} 5393}
\bibitem{Clifford}{de Crombrugghe M. and Rittenberg V. 1983 {\em Ann. Phys.}
{\bf 151} 99}
\bibitem{paper2} {Bilstein U. and Wehefritz B., {\it The XX--model with 
boundaries -- Part II: Magnetization profiles and boundary states}, 
the following paper}
\bibitem{paper3} {Bilstein U. and Wehefritz B., {\it The XX--model with
boundaries -- Part III: The magnetic charge operator for finite chains 
and its role in the continuum limit}}
\end{thebibliography}
\end{document}